\documentclass[]{aa}

\usepackage{graphicx, multirow}
\usepackage{txfonts}
\usepackage{hyperref}
\begin{document}

   \title{Machine Learning Based Classification of Type 1 Active Galactic Nuclei}

   \author{Jie Li\inst{1,2}
          \and
          Xu-Liang Fan \inst{1,2}
          }

    \institute{School of Mathematics, Physics and Statistics, Shanghai University of Engineering Science, Shanghai 201620, People's Republic of China\\
      \email{fanxl@sues.edu.cn}
      \and
     Center of Application and Research of Computational Physics, Shanghai University of Engineering Science, Shanghai 201620, People's Republic of China\\
             }

   \date{Received xxx; accepted xxx}


  \abstract
   {Quasars and Seyfert galaxies are two prototypes of type 1 AGNs. Current view believes that they belong to a uniform population with continuous distributions of bolometric luminosity and host galaxy contribution. Clarifying whether physical differences exist between them would be helpful for understanding the nature of accretion of supermassive black hole and the co-evolution with their host galaxies. Although the sample size and researches of quasars are growing rapidly recent years, researches on Seyfert galaxies are yet limited. Methods to select Seyfert galaxies in large surveys, or distinguish quasars and Seyfert galaxies in broad-line AGNs also receive little attention in the literature.}
   {We aim to provide a machine learning based method to distinguish type 1 quasars and type 1 Seyfert galaxies based on photometric data, which could be applied to large all-sky surveys. The method can also be used to obtain relatively complete samples of Seyfert galaxy candidates when broad-line AGNs have been selected.}
   {Principle component analysis (PCA) is firstly applied to reduce the dimensionality and visualize the separation between quasars and Seyfert galaxies based on multidimensional data. Then we train an supervised classifier to distinguish quasars and Seyfert galaxies with optical and mid-infrared photometric data.}
   {The results of PCA show that quasars and Seyfert galaxies are partial distinguishable. The supervised classifier based on random forest (RF) achieves good performances for type 1 quasars with a F$_1$ score of 0.967. The performance for classifying Seyfert galaxies is worse, with a F$_1$ score of 0.791. Increasing the classification threshold can help to derive a sample of Seyfert galaxies with high purity. The classifier are then applied to the cross-matched sample with DESI DR1. When compared with the results of spectroscopic classification, quasars also show high precision and recall. For Seyfert galaxies identified by RF classifier, there are plenty of less luminous sources who are classified as quasars by spectroscopic method. If they are regarded as Seyfert galaxies, our classifier also performs well to predict Seyfert galaxies. Both the median SEDs and the fraction of Seyfert galaxies derived by our classifier are similar with those for the training sample, while the fraction of Seyfert galaxies for spectroscopic classification are lower at all redshift. These results indicate that our classifier can recover the features of Seyfert galaxies, at least those similar to the Seyfert galaxies in the training sample.}
   {}


\titlerunning{Type I AGN Classification via ML}
\authorrunning{Li \& Fan}
   \maketitle
   \nolinenumbers
%

\section{Introduction} \label{sec:intro}
Active galactic nuclei (AGNs) are compact regions at the center of galaxies, whose energy can exceed the total energy released by their host galaxies. AGNs can emit across the entire electromagnetic spectrum, which makes them important targets for multi-wavelength observations~\citep{1997ARA&A..35..445U, 2006ApJS..166..470R, 2010ApJ...716...30A, 2010Natur.463..919A}. However, depending on the characters at different wavelengths, AGNs can be classified as various types~\citep{2015ARA&A..53..365N, 2017A&ARv..25....2P}. Type 1 AGNs, whose optical spectra show both broad and narrow emission lines, are the predominant type of them. Type 1 Seyfert galaxies and type 1 quasars\footnote{For simplicity, we omit type 1 in the following text.} are two typical types of type 1 AGNs. Historically, Seyfert galaxies were defined by their extended galaxy morphologies and strong emission lines~\citep{1943ApJ....97...28S}. Quasars were firstly discovered by radio surveys. Their optical counterparts were found to locate at relatively high redshift~\citep{1963Natur.197.1040S}. The radiation from quasars is overwhelming that from host galaxies, which results in their point-like optical morphologies. As the samples growing, a population of Seyfert galaxies with relatively smaller width of emission lines are found, which are named as narrow line Seyfert 1 galaxies (NLS1s, \citealt{1985ApJ...297..166O, 2001A&A...372..730V}). Compared with the classic broad line Seyfert 1 galaxies (BLS1s), NLS1s show higher accretion rate and lower black hole mass~\citep[e.g.,][]{2012AJ....143...83X}. Among quasars, there are also a population with red spectrum energy distributions (SEDs), named red quasars~\citep{1995Natur.375..469W, 2003AJ....126.1131R}. Their properties are also found to be different with classic quasars~\citep[e.g.,][]{2019MNRAS.488.3109K}. In addition, different shapes of SEDs are also found among type 1 AGNs~\citep{2014MNRAS.438.2253S, 2019MNRAS.489.3351B, 2023NatAs...7.1506C}. 

The optical spectroscopic properties of Seyfert galaxies and quasars are similar. Both of them show strong emission lines, while the contribution from their hosts could be obvious for Seyfert galaxies. This results in the difference on the SEDs of them. The optical/UV SEDs of quasars are dominated by a big blue bump from accretion disk, while emission of host galaxies could dominate the optical SEDs of Seyfert galaxies. 

Another difference between quasars and Seyfert galaxies are their luminosity. The luminosity of Seyfert galaxies is generally lower than that of quasars, although a continuous sequence in luminosity seems to exist between these two classes. The physical driver of the luminosity difference is still unclear. Potential factors include the central black hole mass, accretion process, and (or) the host galaxies of them. Quasars are believed to dominate the high end of black hole mass function~\citep{2007ApJ...667..131G, 2010ApJ...719.1315K}. The black hole mass of quasars generally ranges from $10^{7}$ to $10^{11}$ solar mass~\citep{2011ApJS..194...45S, 2022ApJS..263...42W}, with the peak at $10^{7}$ to $10^{9}$ solar mass for different redshift bins~\citep{2009ApJ...699..800V}. Seyfert galaxies host lighter black hole than quasars.~\citet{2012AJ....143...83X} explored a sample of broad line Seyfert 1 galaxies, and found their black hole mass ranges from 10$^{6.5}$ to 10$^{8.4}$ solar mass with an average value of 10$^{7.2}$ solar mass. For the Seyfert galaxy sample from SDSS DR17, the average black hole mass is 10$^{7.9}$ solar mass~\citep{2024MNRAS.527.7055P}\footnote{Note that these Seyfert galaxies samples also contain sources with high luminosity, i.e., quasars~\citep{2003ApJ...585..647B, 2024MNRAS.527.7055P}.}.

\begin{figure}[ht!]
    \centering
    \includegraphics[width=0.45\textwidth]{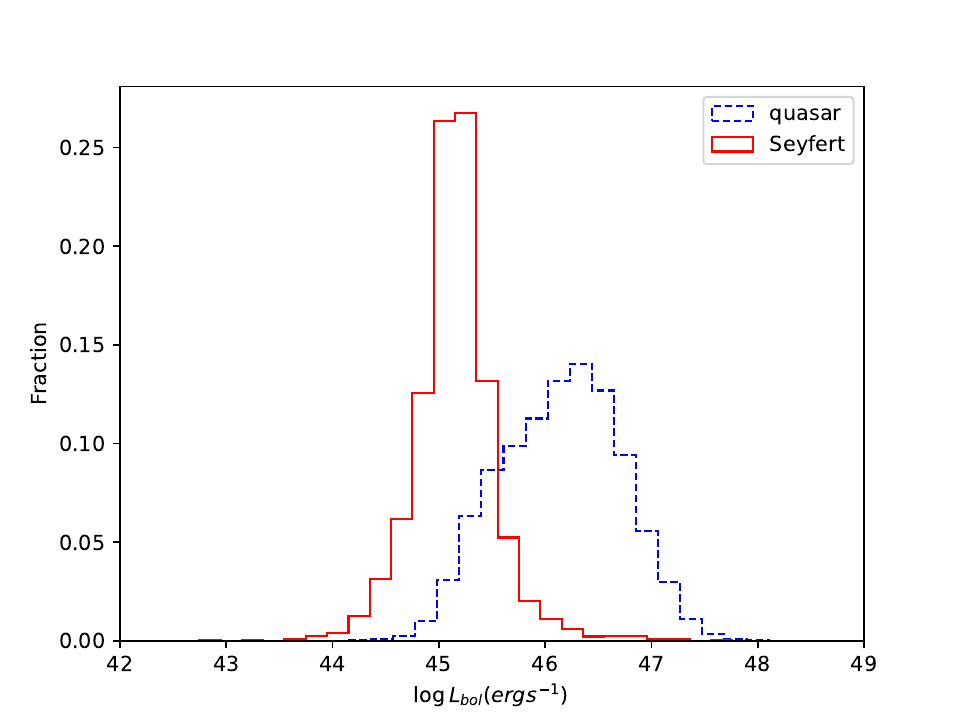}
    \includegraphics[width=0.45\textwidth]{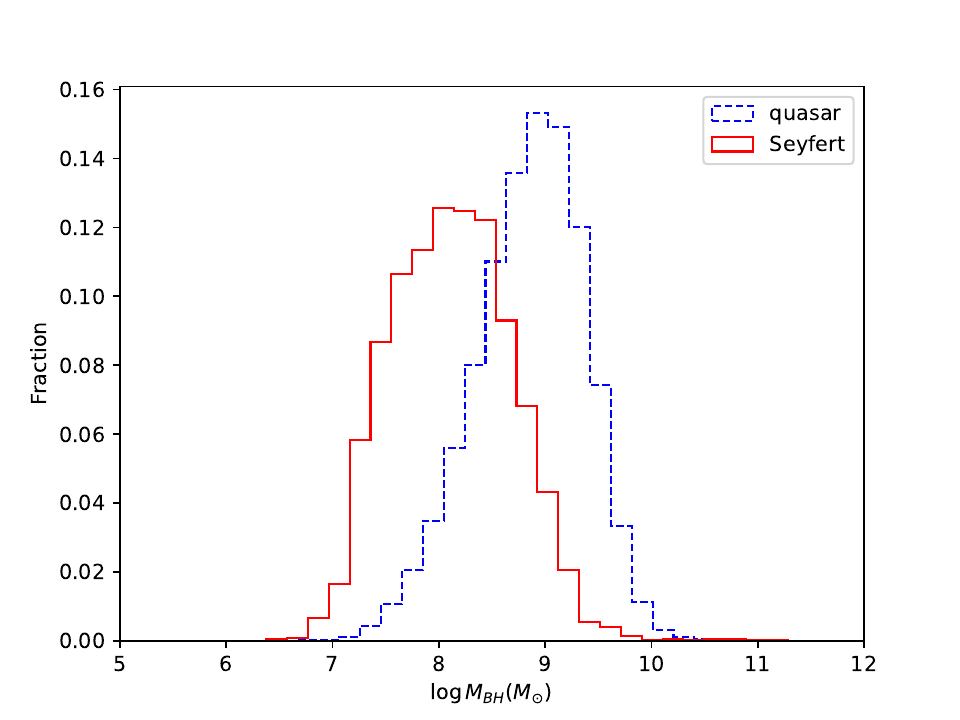}
    \includegraphics[width=0.45\textwidth]{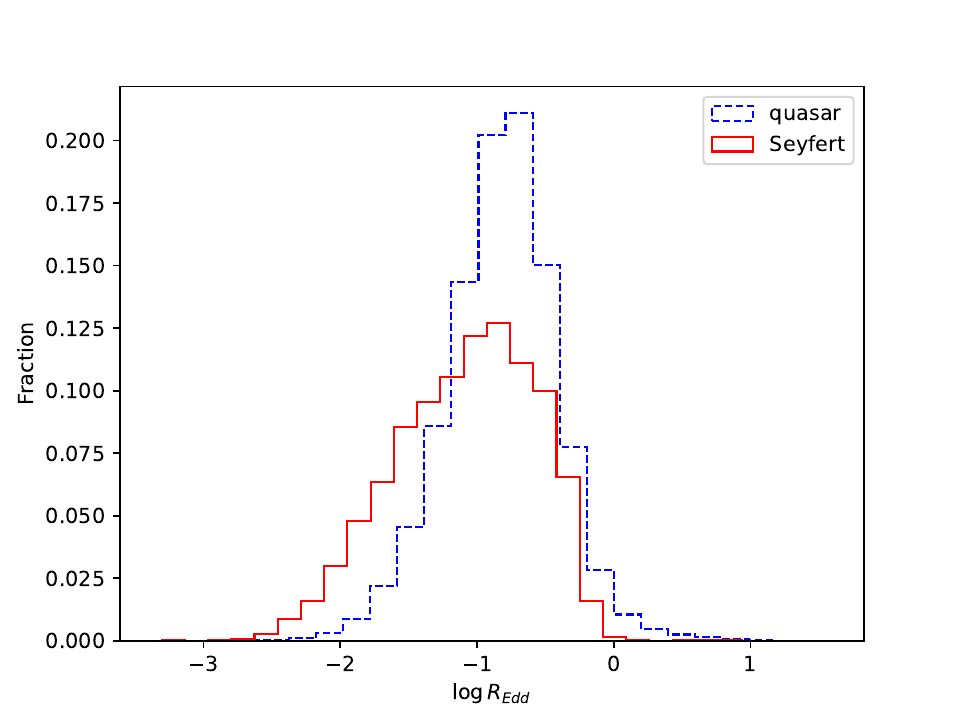}
    \caption{From top to bottom, distributions of bolometric luminosity, black hole mass, and Eddington ratio for quasars and Seyfert galaxies are shown, respectively. }
    \label{fig:phy_par}
\end{figure}

Different accretion modes are found in Seyfert galaxies and quasars, e.g., advection
dominated accretion flows, thin disk accretion, and super-Eddington accretion~\citep{1999ApJ...526L...5L, 2002A&A...388..771C}. The Eddington ratio of broad line Seyfert 1 galaxies range from 0.04 to 0.5 (\citealt{2012AJ....143...83X}, the distribution is slightly broader for the sample of SDSS DR17, \citealt{2024MNRAS.527.7055P}), while the bolometric luminosity can approach or even exceed the Eddington luminosity in quasars (see a recent review of~\citealt{2025Univ...11...69M}, also see Figure~\ref{fig:phy_par} for the distributions of bolometric luminosity, black hole mass, and Eddington ratio in this work).

Quasars are found to reside in massive elliptical galaxies by early observations~\citep{2003MNRAS.340.1095D, 2004MNRAS.355..196F}, while the host galaxies of Seyfert galaxies tend to be spiral galaxies~\citep{1977ApJS...33...19A, 1998ApJS..117...25M, 2017Ap&SS.362..230C}. Lately, disk-like structures were also found in the host galaxies of quasars~\citep{2014MNRAS.440..476F, 2022ApJ...934..130Z, 2023Natur.621...51D}. The fraction of late-type hosts can be up to 40 percent~\citep{2022ApJ...934..130Z}. In addition, redshift distributions of quasars and Seyfert galaxies are also different. Quasars are discovered at various redshift, from several hundred million years after the Big Bang~\citep{2011Natur.474..616M, 2015Natur.518..512W, 2018Natur.553..473B} to the local universe~\citep{2011ApJS..194...45S, 2022ApJS..263...42W}. Seyfert galaxies are usually discovered at low redshift. One possible reason for this is that high-redshift Seyfert galaxies are hard to be detected due to their low luminosity.

Although various differences exist between quasars and Seyfert galaxies, they are believed to belong to a unified population in the literature~\citep{2015ARA&A..53..365N, 2018ARA&A..56..625H}, where their luminosity, fraction of host contribution, and other features form continuous distributions. Furthermore, there are also suggestions that SED, host galaxy environment, and cosmological evolution of AGNs are dependent on luminosity~\citep{2005A&A...441..417H, 2006ApJS..166..470R, 2022A&A...663A.130M}. Therefore, clarifying whether physical differences exist between quasars and Seyfert galaxies is helpful for understanding the nature of accretion of supermassive black hole and the co-evolution with their host galaxies. However, Seyfert galaxies, as a prototype of AGNs, their population studies have received less attention in recent years. In recent optical surveys, Seyfert galaxies are even not defined as an individual spectral type. This is partly due to that Seyfert galaxies are difficult to be separated with quasars in large-area surveys. Historically, quasars and Seyfert galaxies are usually separated based on their absolute magnitude. The critical B band absolute magnitude is taken as $-23.0$ or $-21.5 + 5\log h_0$~\citep{1983ApJ...269..352S, 2010A&A...518A..10V}. This criterion just focus on the difference on the luminosity of type 1 AGNs, but no information about the host galaxy contribution is considered. \citet{2015PASA...32...10F} applied a criterion based on the optical extended structures. However, for faint AGNs or AGNs with large redshift, extended features from host galaxies are difficult to constrain. Seyfert galaxies can also be selected in spectroscopic surveys. The spectroscopic objects are normally classified as quasars, galaxies, and stars by template fitting method~\citep{2012AJ....144..144B}. Then AGNs in galaxies can be identified through the so-called BPT diagram~\citep{1981PASP...93....5B}. These AGNs with broad emission lines can be regarded as Seyfert galaxies, as they own the emission line properties of AGNs and potential contributions from host galaxies. This method still relies on the separation of host and AGN components in optical spectra, as well as a robust estimation of redshift. Whether this criterion can yield a complete sample of Seyfert galaxies has yet to be explored.

Machine learning based classification has received more and more attentions in recent years. It is widely used in the tasks of separating quasars from galaxies and stars~\citep{2012MNRAS.425.2599P, 2018arXiv180809955B, Jin_2019, 2020A&A...633A.154L, Li_2021, 2024A&A...690A.224L, 2024ApJS..271...54F, 2025ApJS..279...26F}. This task can be achieved based on the optical/infrared photometric~\citep{2020A&A...633A.154L, 2024ApJS..271...54F}, optically spectroscopic~\citep{2018arXiv180809955B, 2024MNRAS.533.1976M} and(or) imaging features~\citep{2023MNRAS.524.5080S, 2025ApJS..279...26F}. These machine learning based classifiers have been used to build samples of quasar candidates for large-area surveys~\citep{2018arXiv180809955B, 2024ApJS..271...54F, 2025ApJS..279...26F}. Machine learning is also applied to distinguish different subclasses of AGNs~\citep{Cavuoti_2013, Chang_2021, Chen_2021, 2024MNRAS.527.1163W, 2024MNRAS.527.4677Z}. Experiments have been performed to classify AGNs and non-AGNs~\citep{Cavuoti_2013, 2024MNRAS.527.1163W, 2025A&A...701A.179F}, AGNs selected at different bands~\citep{Chang_2021}, Seyfert galaxies and low-ionization nuclear emission-line regions~\citep{Cavuoti_2013}, type 1 and type 2 Seyfert galaxies~\citep{Cavuoti_2013, Chen_2021}, morphologies of radio galaxies~\citep{2017ApJS..230...20A, 2019ApJS..240...34M} and subclasses of blazars~\citep{2019ApJ...872..189K, 2020ApJ...895..133X}. Owing to the difference on their SEDs, machine learning algorithms are also expected to be efficient to separate quasars and Seyfert galaxies. However, few attempts has been performed to distinguish them in the literature. \citet{2024RNAAS...8...46A} attempted to classify quasars and Seyfert galaxies with SDSS optical spectra based on machine learning. They found much worse performance on Seyfert galaxies when compared to quasars. They also proposed that Seyfert galaxies with high redshift was hard to distinguish from quasars, which indicated that their classifier just revealed the difference on redshift between quasars and Seyfert galaxies~\citep{2024RNAAS...8...46A}.

In this paper, we provide a machine learning based method to identify type 1 Seyfert galaxies and type 1 quasars with photometric data at optical and MIR bands, based on a specific classification method of these two types of type 1 AGNs. Section \ref{sec:methods} describes the machine learning algorithm and metrics used in this paper. Section \ref{sec:data} describes the datasets where the reference labels\footnote{The reference labels throughout this paper only refer to the sources following one specific classification method, rather than the intrinsic physical types of sources.} and the training features are derived. In section \ref{sec:RESULT}, the results of the unsupervised and supervised learning, as well as the application of our supervised classifier, are presented. Summaries can be found in section \ref{sec:summary}. A $\Lambda$CDM cosmology model with $\Omega_{\lambda} = 0.7$, $\Omega_{m} = 0.3$ and $H_{0} = 70~$km s$^{-1} $Mpc$^{-1}$ is used in our calculations. Throughout the paper, MIR flux densities are converted from magnitude with zero-magnitude flux density of 309.540 Jy, 171.787 Jy, 31.674 Jy, and 8.363 Jy for W1, W2, W3, and W4 bands, respectively~\citep{2011ApJ...735..112J}. The AB magnitudes in PS1 DR1 are converted into flux densities with 3631 Jy~\citep{2012ApJ...750...99T}. The effective wavelength for each band is also taken from~\citet[their Table 4]{2012ApJ...750...99T}. The galactic extinction is corrected with the dust map of ~\citet{1998ApJ...500..525S} and extinction curve of~\citet{1999PASP..111...63F}. The k-correction is performed with a power-law spectral index $\alpha = 1$, which is derived by fitting the median SEDs of the flux density for the main sample (Section~\ref{sec:data})\footnote{The spectral indexes for quasars and Seyfert galaxies are $1.01\pm0.05$ and $1.06\pm0.06$, respectively.}.

\section{Methods} \label{sec:methods}
Machine learning can be generally divided into unsupervised and supervised learning. Both approaches have been widely applied to the classification of astronomical objects~\citep{2016ApJS..225...31L, 2017ApJS..230...20A, 2018MNRAS.476.3661D, 2018MNRAS.473.1108H, 2020A&A...633A.154L}. Unsupervised machine learning algorithms, including clustering and dimensionality reduction, need no pre-defined labels about the source class. They can be used to explore and visualize new characters from the datasets. Principle component analysis (PCA) is one of the widest applied algorithm to astronomical data, which can help to clarify the physical drivers of observational differences and determine the optimal features for supervised learning~\citep{1992ApJS...80..109B, 2020A&A...633A.154L, 2020ApJ...895..133X}. PCA constructs a covariance matrix of the dataset and computes its
eigenvectors~\citep{2019arXiv190407248B}. It can be used as a dimensionality reduction technique, which transform high-dimensional data into a lower-dimensional space. The new coordinate system is created by projecting the original data aligned with the direction of the maximum variance. In this paper, PCA is applied to reduce the dimensionality and visually verify whether quasar and Seyfert galaxies are separable when specific combinations of features are chosen.

The use of supervised machine learning is more frequent to construct the classier of different astronomical objects~\citep{2016ApJS..225...31L, 2017ApJS..230...20A, 2018MNRAS.476.3661D, 2025A&A...701A.179F}. The biggest difference of supervised learning compared with unsupervised learning is that the former relies on the training phase based on the known classification labels. There are various supervised learning algorithms in practical applications~\citep{2019arXiv190407248B}. Random forest (RF, \citealt{2001MachL..45....5B}) and XGBoost (XGB, \citealt{2016arXiv160302754C}) are chosen in this study to classify type 1 AGNs. RF and XGB are two of ensemble algorithms, which can improve the model performance by combining the information of base learners. RF is one of the widely and early used algorithm in astronomical data~\citep{2010ApJ...712..511C, 2011ApJ...733...10R}. It achieves the ensemble by constructing a collection of decision trees. For different decision trees, the features and data subsets are randomly selected in the training process~\citep{2001MachL..45....5B}. XGB, as an optimized gradient boosting ensemble method, is recently applied to the task of classification of astronomical objects, and show good performance for large datasets~\citep{2021ApJS..254....6F, Li_2021, 2025ApJS..279...26F}.

As shown in section~\ref{sec:data}, the number of quasars and Seyfert galaxies in the training sample is imbalance, the performance of the machine learning could be dominant by the majority class. There are many approaches to overcome this problem. we employ the Synthetic Minority Oversampling Technique (SMOTE) in this work~\citep{chawla2002smote}, which balance the sample of two classes by over-sampling the minority class.

In order to evaluate the performance of the classifier on both classes, the precision, recall and $F_1$ score of each class are calculated separately. Precision and recall can evaluate the ability of the classifier to find positive samples without misclassification and find all the positive samples, respectively. $F_1$ score is the harmonic mean of precision and recall. The three metrics are defined as follows:
\begin{equation}
    precision = \frac {TP} {TP+FP}
\end{equation}
\begin{equation}
    recall = \frac {TP} {TP+FN}
\end{equation}
\begin{equation}
    F_1 = 2 \times \frac {precision \times recall} {precision+recall}
\end{equation}
where TP represents the number of positive samples that are correctly identified. FP is the count of positive samples that are incorrectly identified. FN is the count of negative samples that are incorrectly classified.

In this paper, the learning algorithms are running with the Python libraries \textit{scikit-learn} \citep{2011JMLR...12.2825P} and \textit{xgboost}\citep{2016arXiv160302754C}. All the metrics are evaluated through the test set, which is derived from the randomly split of the training data. 70\% of the training data are used to train the model (training set), while remaining 30\% are taken as test set.

\section{Properties of Training Data} \label{sec:data}
To perform a supervised learning, samples with known classification are needed. As mentioned in Section~\ref{sec:intro}, all current classification methods of quasars and Seyfert galaxies have limitations. A well-performing machine learning classifier should reproduce the source properties of the training sample. But it cannot identify which classification method reflects the intrinsic physical distinction between quasars and Seyfert galaxies. In this work, we choose the classification in the million quasar catalog (Milliquas, \citealt{2023OJAp....6E..49F}) as the reference labels for machine learning. Milliquas compiled published AGN catalogs in the literature before June 30, 2023. Majority of type 1 AGNs in it come from the spectroscopic observations of Sloan Digital Sky Survey (SDSS, \citealt{2020ApJS..250....8L}). For its eighth version, Milliquas contains more than one million AGNs (or AGN candidates) of various types, including 860100 type 1 quasars (\texttt{TYPE = `Q'}) and 47044 type 1 Seyfert galaxies (\texttt{TYPE = `A'},~\citealt{2023OJAp....6E..49F}). Seyfert galaxies and quasars are firstly compiled from the spectroscopic catalogs. Then they are distinguished by identifying whether extended morphologies exist or not on their optical images. For faint sources without constraint on optical morphology, quasars and Seyfert galaxies are separated by defining a critical value of redshift corrected magnitude~\citep{2015PASA...32...10F}, which is similar to that based on the absolute magnitude~\citep{2010A&A...518A..10V}. Sources brighter than this critical value are classified as quasars, while the faint sources are Seyfert galaxie~\citep{2015PASA...32...10F}. The comparison between samples derived from this method with that from spectroscopic method will be discussed in Section~\ref{sec:Application}. 

Optical and MIR photometric data are widely used to classify AGNs~\citep{1997iagn.book.....P, 2006ApJS..166..470R, 2012ApJ...753...30S, 2018ApJS..234...23A}. For these two bands, differences between quasars and Seyfert galaxies can come from the colors of their host galaxies, the ratio between AGN and host components, as well as redshift. Therefore, photometric measurements at optical and MIR bands are expected to be effective for distinguishing these two types of AGNs. At mid-infrared (MIR) band, Wide-field Infrared Survey Explorer(WISE, \citealt{2010AJ....140.1868W}) conducted an all-sky survey using four filters with the central wavelength at 3.4, 4.6, 12, and 22 $\mu$m (hereafter $W1$, $W2$, $W3$ and $W4$), respectively. The AllWISE catalog provided magnitude measurements of more than 747 million sources~\citep{2014yCat.2328....0C}. The Panoramic Survey Telescope and Rapid Response System(Pan-STARRS) is a wide-field optical/NIR imaging system. Its first stage (PS1) uses a 1.8-meter telescope and a 1.4 Gigapixel camera to image the sky in five broadband filters (g, r, i, z, y, \citealt{2010SPIE.7733E..0EK}). The photometric catalog of PS1 DR1 contains more than 1.9 billion objects~\citep{2016arXiv161205560C, 2020ApJS..251....6M}. The photometric data from AllWISE and PS1 are used to construct the features for machine learning.

\begin{figure}[ht!]
    \centering
    \includegraphics[width=0.45\textwidth]{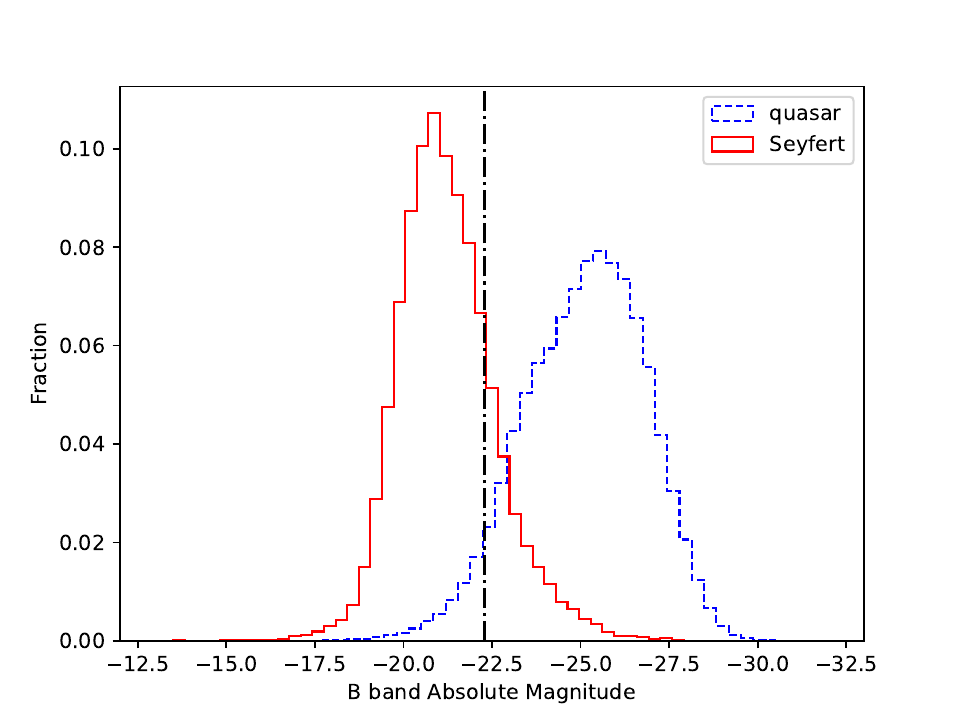}
    \includegraphics[width=0.45\textwidth]{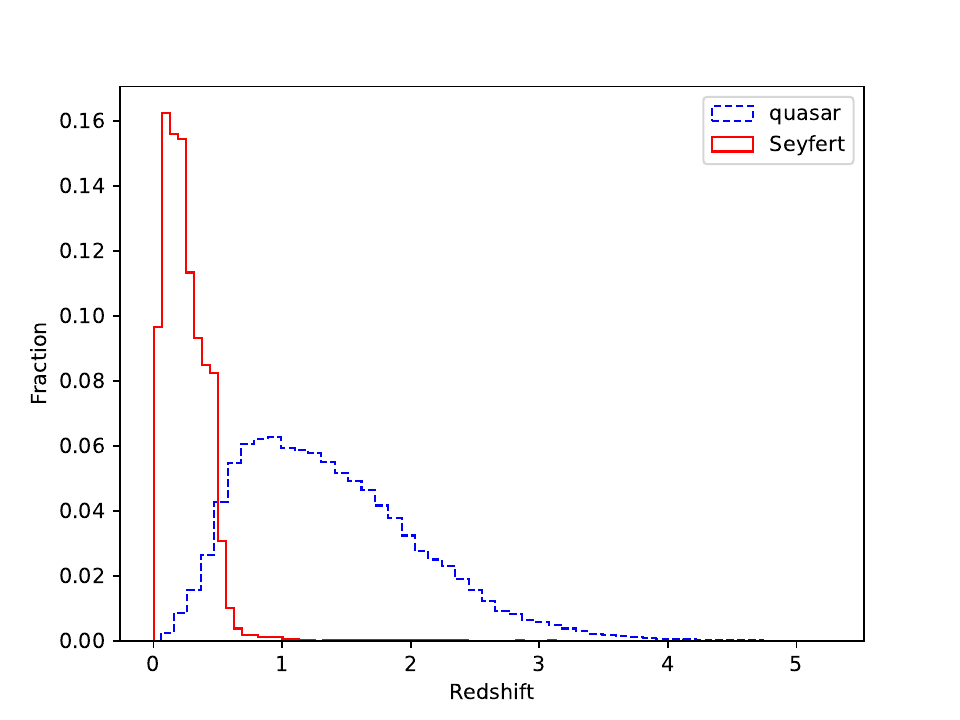}
    \caption{Top panel: Distributions of B band absolute magnitude for quasars and Seyfert galaxies. The black line represents the critical value ($-22.3$) for distinguishing quasars from Seyfert galaxies. Bottom panel: Distributions of redshift.}
    \label{fig:sample_pro}
\end{figure}

TOPCAT \citep{2005ASPC..347...29T} is applied to cross-match the catalog of AllWISE, PS1 with Milliquas, with a matching radius of 1\arcsec. The sources with multiple matches are removed to avoid spurious matches. After removing the sources without magnitude measurements and their corresponding errors at each of nine filters of WISE and PS1, 150,373 quasars and 24,881 Seyfert galaxies are left. These sources make up the main sample for unsupervised and supervised machine learning. Figure~\ref{fig:sample_pro} plots the distributions of B band absolute magnitude\footnote{The absolute magnitude is calculated with $M = m - DM - k(z)$, where $DM$ is the distance modulus, $k(z) = 2.5(\alpha-1)\log(1+z)$ is the term of k-correction~\citep{1997iagn.book.....P}.} and redshift for quasars and Seyfert galaxies of the main sample. Quasar and Seyfert galaxies show obvious discrepancy in both plots. The critical absolute magnitude ($-22.3$\footnote{The divider is derived with $-21.5+5\log h_0$ according to the cosmology model we used~\citep{1997iagn.book.....P, 2010A&A...518A..10V}.}) to separate quasars and Seyfert galaxies is also plotted in the top panel of Figure~\ref{fig:sample_pro}. The distribution of absolute magnitude is bimodal. Generally, quasars show absolute magnitude brighter than $-22.3$ mag, while Seyfert galaxies is opposite. For redshift, Seyfert galaxies generally have redshift less than 1, while distribution of the redshift for quasars are very broad.

\begin{figure*}[ht!]
    \centering
    \includegraphics[width=0.9\textwidth]{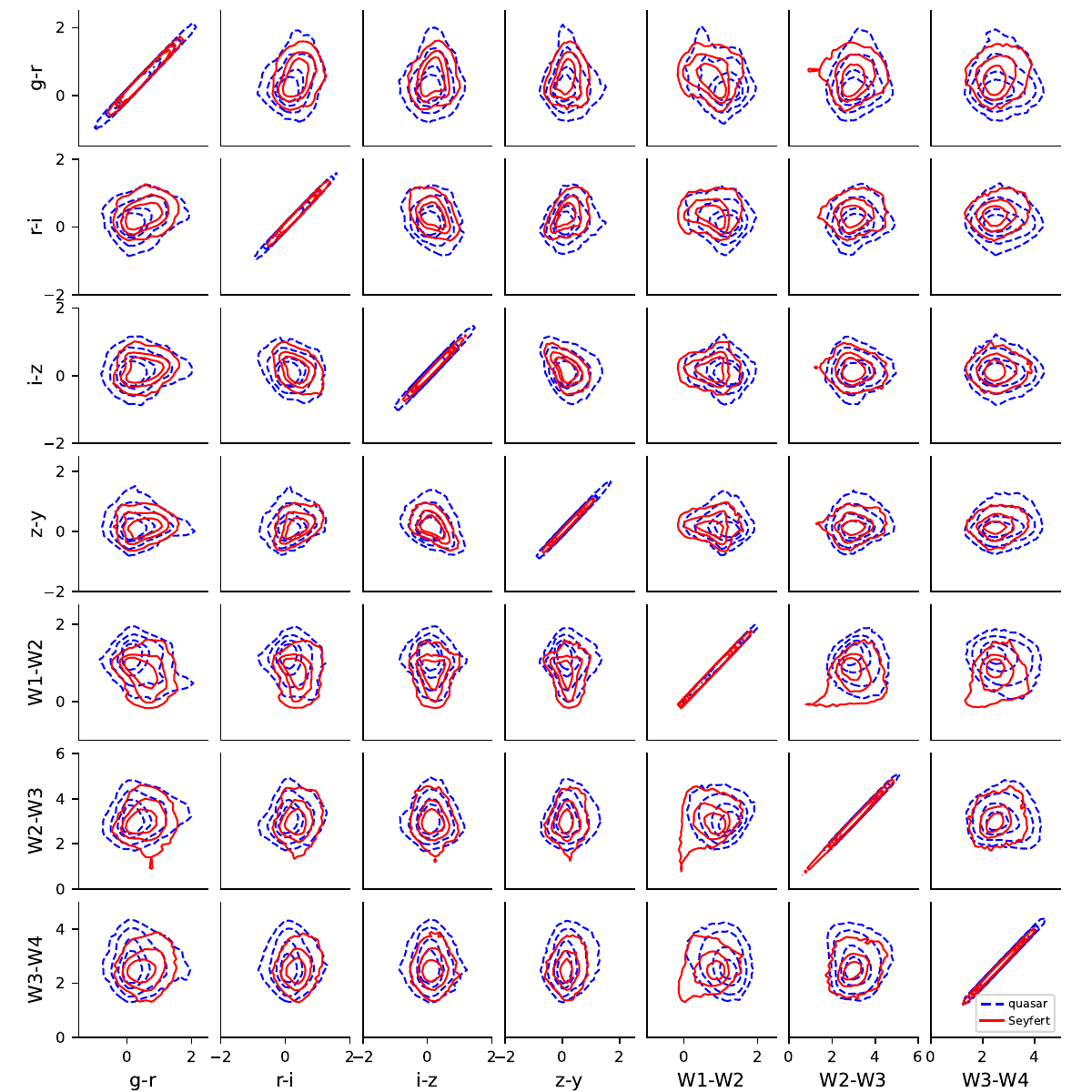}
    \caption{Optical and MIR color-color diagrams of quasars (blue contours) and Seyfert galaxies(red contours). The levels of blue contours are [10, 100, 1000, 3000], while the levels of red contours are [10, 100, 300]. }
    \label{fig:color}
\end{figure*}

The main sample is then cross-matched with SDSS DR16~\citep{2022ApJS..263...42W} to compare several physical properties for quasars and Seyfert galaxies. This results in a sample of 125,089 quasars and 6,840 Seyfert galaxies. The distributions of bolometric luminosity, black hole mass, and Eddington ratio for these two types of sources are plotted in Figure~\ref{fig:phy_par}. In general, quasars show higher bolometric luminosity and black hole mass than Seyfert galaxies. The difference on Eddington ratio is small, except that Seyfert galaxies show a tail at lower Eddington ratio.

Figure~\ref{fig:color} shows the color-color diagrams for optical and MIR colors. For most colors, quasars and seyfert galaxies are indistinguishable. The centroids are slightly misaligned for three colors, i.e., W1-W2, g-r, and r-i. Seyfert galaxies show bluer W1-W2 color than that of quasars. For g-r and r-i, Seyfert galaxies are slightly redder than quasars. Distinguishing quasars and Seyfert galaxies is yet difficult through comparing the differences of color pairs. We employ PCA algorithm to explore whether the multi-dimensional features could enhance the separation.

\section{Results and Disscussions} \label{sec:RESULT}
\subsection{Results of principal component analysis\label{subsec:pca}}
\begin{figure}[ht!]
    \centering
    \includegraphics[width=0.45\textwidth]{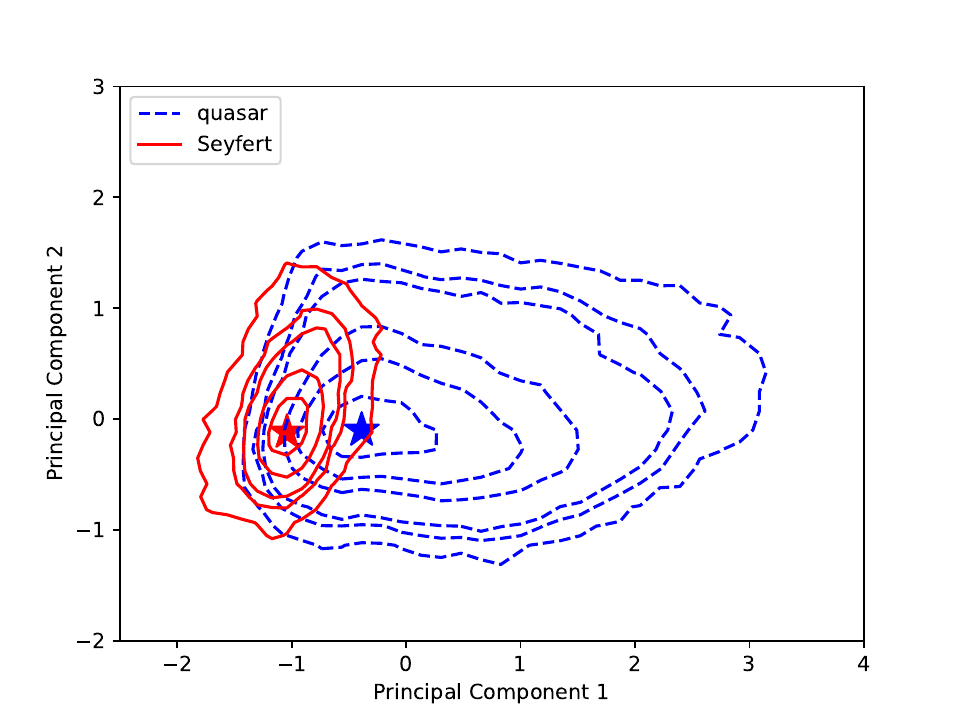}
    \includegraphics[width=0.53\textwidth]{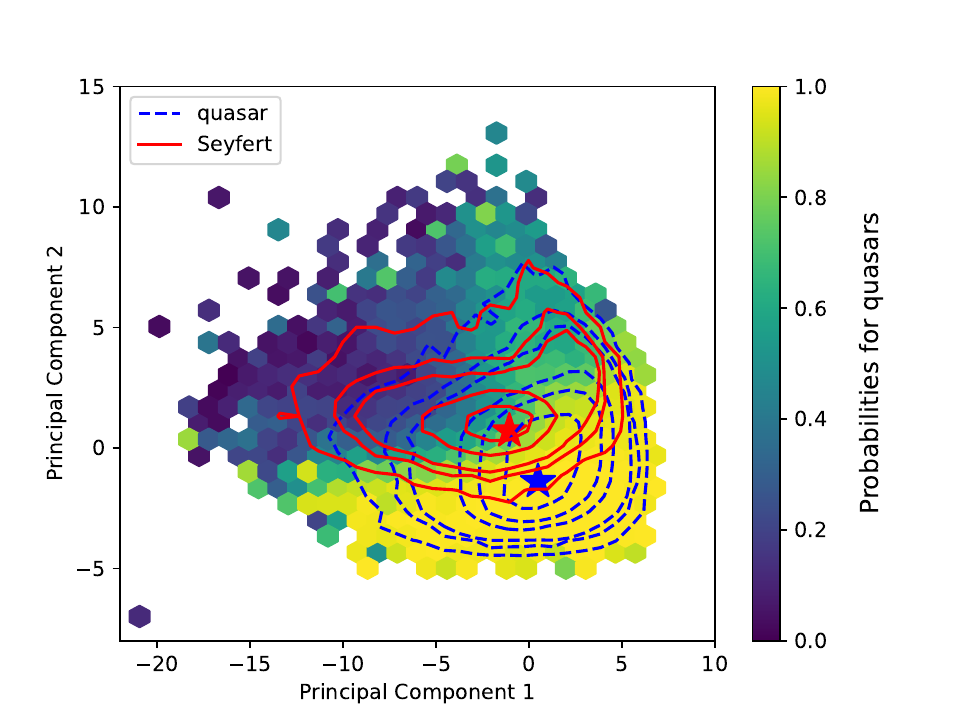}
    \caption{
        Results of dimensionality reduction with PCA. Top panel is derived with the feature combination of redshift, optical and MIR colors. Bottom panel is derived with the feature combination of optical and MIR magnitudes and colors, as well as the optical/MIR colors (see the text for details). The supervised learning is also performed with this feature combination. The classification from the RF based classifier are overplotted as the hexbin. The mean value of the predicted probabilities for quasars is calculated in each hexagonal bin, which is also represented by the color bar. The red and blue stars in each panel show the locations of highest number densities for Seyfert galaxies and quasars, respectively. The levels of quasar contours are [10, 50, 100, 500, 1000, 2000], while the levels of Seyfert galaxy contours are [10, 50, 100, 300, 600].
    }
    \label{fig:pca}
\end{figure}

For visual inspection, PCA is applied to reduce the dimensionality to two dimensions. Two combinations of features are shown to clarify whether quasars and Seyfert galaxies are distinguishable. One is redshift combining with the optical and MIR colors (W1-W2, W2-W3, W3-W4, g-r, r-i, i-z, z-y). The other is the optical and MIR magnitudes and colors (W1, W2, W3, W4, W1-W2, W2-W3, W3-W4, g, r, i, z, y, g-r, r-i, i-z, z-y), combining with the colors between optical and MIR (g-W1, r-W1, i-W1, z-W1, y-W1). Figure~\ref{fig:pca} shows the results after the dimensionality reduction. The top panel shows the result for the feature group including redshift, while the bottom panel is derived from the feature group including magnitude measurements and optical/MIR colors.

For two combinations of features, quasars and Seyfert galaxies are partially overlapping, while their centroids are misaligned. It is thus expected at least a fraction of the objects in each class could be separated with these features. For the top panel, Principal Component 1 could be dominated by redshift. Quasars at high redshift can be easily separated with Seyfert galaxies. However, low-redshift quasars, who have similar optical and MIR colors with Seyfert galaxies, are difficult to distinguish. PCA analyses with other combinations of features are also explored. When only optical and MIR colors are considered, quasars and Seyfert galaxies almost completely overlap. Adding magnitude measurements yields a slight separation between quasars and Seyfert galaxies. The separation becomes more significant when the colors between optical and MIR are included (bottom panel of Figure~\ref{fig:pca}).

\subsection{Performance of supervised classifier\label{subsec:supervised}}
The results of PCA indicate that quasars and Seyfert galaxies are at least partially distinguishable through two feature groups. Accurate redshift measurement requires the observation of spectroscopy, which is lacking when the supervised learning is performed with photometric data. On the other side, redshift estimation can be approached from photometric observations by machine learning method~\citep{2003MNRAS.339.1195F, 2010ApJ...712..511C, 2022MNRAS.512.4593Z}. Therefore, we prefer to build a supervised classifier without redshift information as an input. The second feature group (W1, W2, W3, W4, W1-W2, W2-W3, W3-W4, g, r, i, z, y, g-r, r-i, i-z, z-y, g-w1, r-W1, i-W1, z-W1, y-W1) are used for the training process. RF is firstly applied to train the model. The classifier achieves a good performance in identifying quasars. The precision, recall and $F_1$ score for quasars can be as high as 0.962, 0.973 and 0.967, respectively. For Seyfert galaxies, the model performance is worse than that for quasars, with $F_1 = 0.791$. The recall (0.762) for Seyfert galaxies gets a lower value than precision (0.822). The detailed metrics are listed in Table~\ref{tab:performance}.

In order to improve the model performance for Seyfert galaxies, SMOTE oversampling technique~\citep{chawla2002smote} is employed to balance the sample size of quasars and Seyfert galaxies. The ratio of minority class to majority class is set to 1. RF with the resampled training set obtains a higher value of recall (0.879) for Seyfert galaxies. However, precision gets worse after the oversampling. The $F_1$ score for Seyfert galaxies also decrease slightly. On the other side, the model performance get worse for quasars, with the $F_1$ score decreases to 0.961 (Table~\ref{tab:performance}).

XGB is also applied for training. Generally, XGB performs similar with RF. The precision for Seyfert galaxies obtained by RF is higher than that by XGB, while it is the opposite for the recall. Combining SMOTE would further improve the performance of recall for Seyfert galaxies. However, the overall performance ($F_1$ score) is worse compared with the performance when only XGB is applied (Table~\ref{tab:performance}). As SMOTE oversampling technique and XGB have no obvious improvement on the model performance, we use RF to build our supervised classifier.

\begin{figure}[ht!]
    \hfill
    \includegraphics[width=0.486\textwidth]{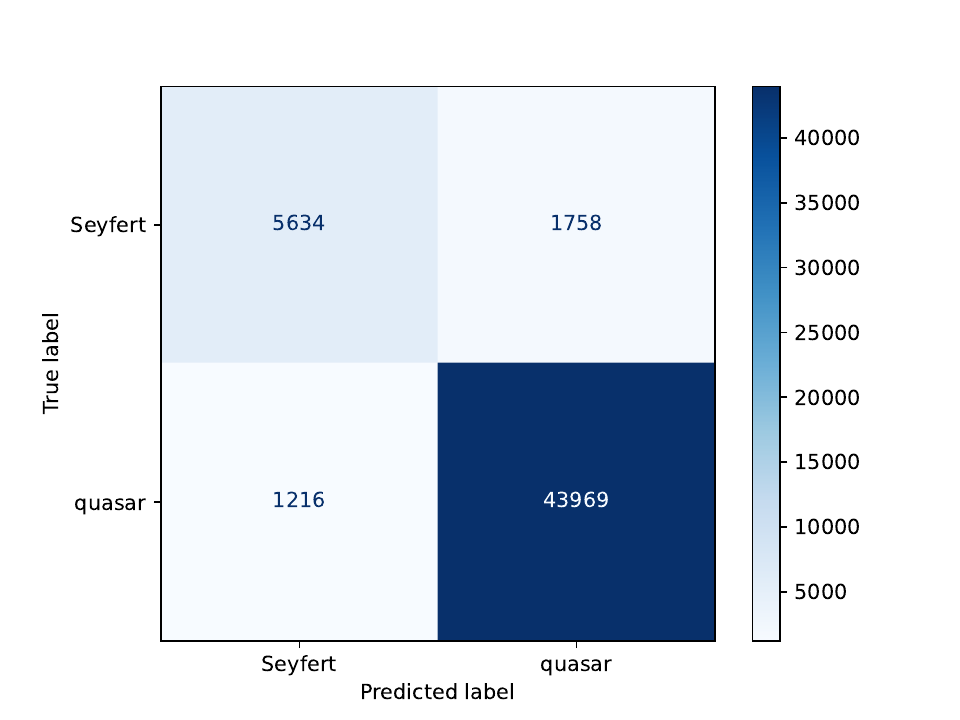}
    \includegraphics[width=0.45\textwidth]{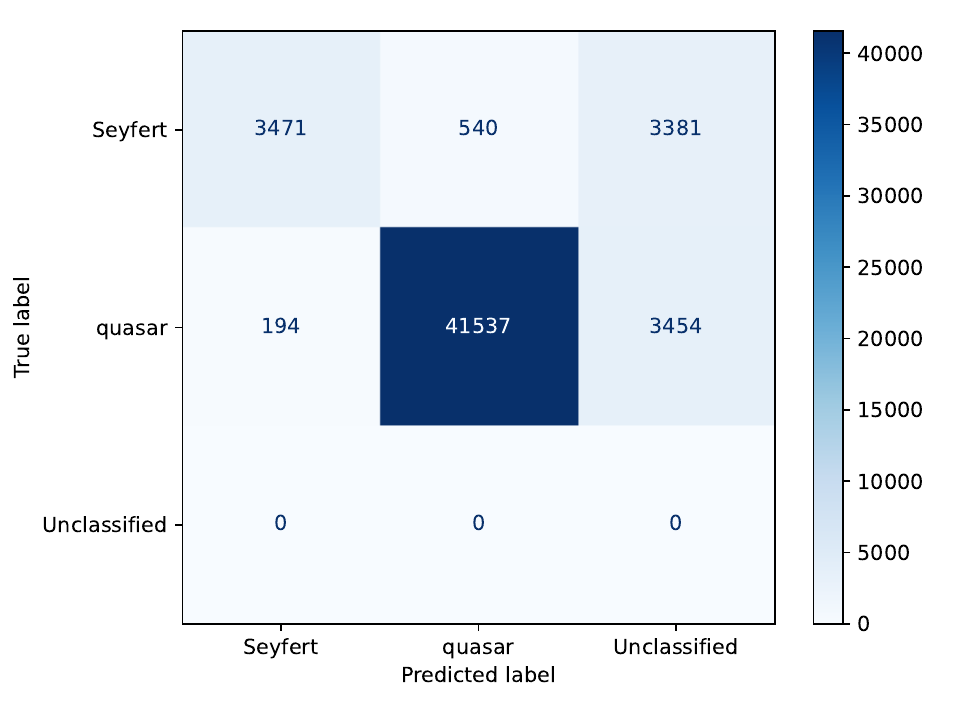}
    \caption{
        Confusion matrix of test set. Top panel is derived when the classification threshold, i.e., the critical predicted probability is 0.5. For the bottom panel, the classification threshold is 0.8.
    }
    \label{fig:matrix}
\end{figure}

\begin{figure}
    \center
    \includegraphics[width=0.5\textwidth]{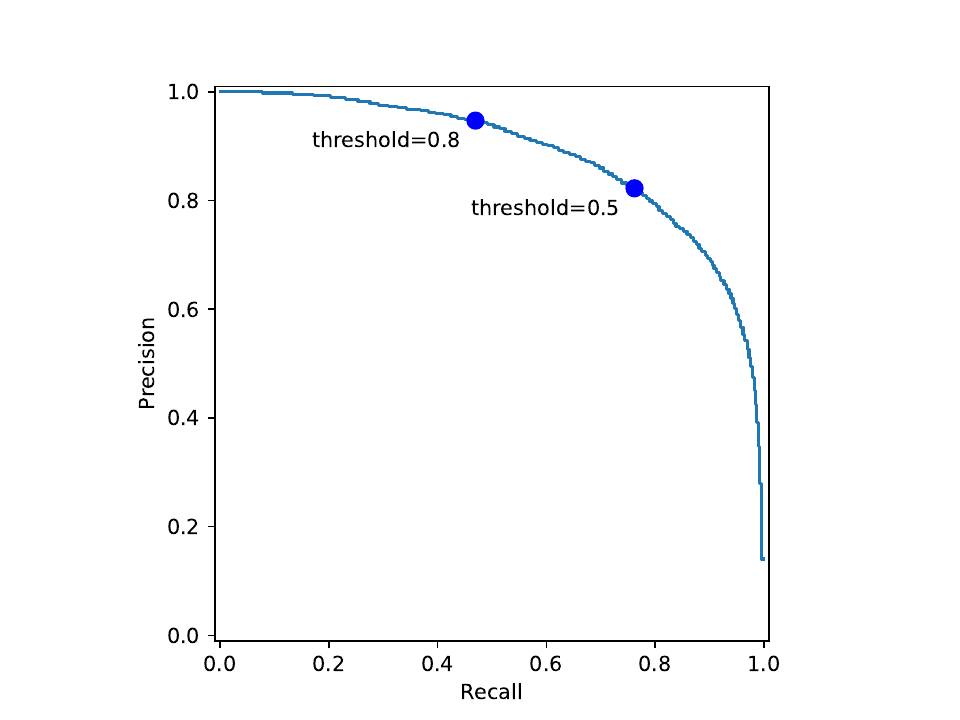}
    \includegraphics[width=0.5\textwidth]{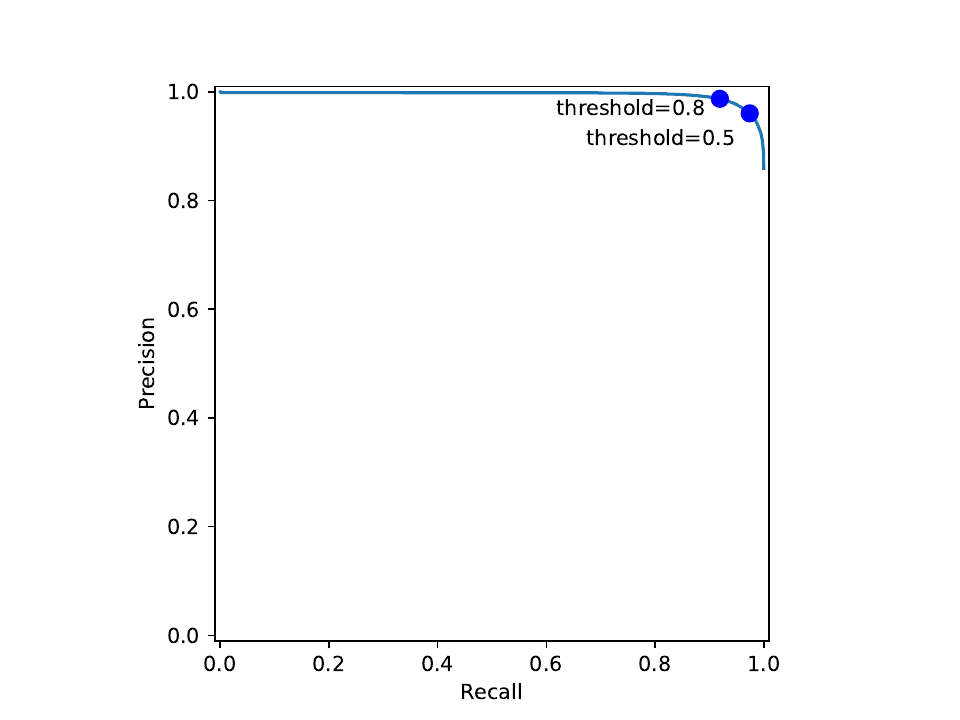}
    \caption{
        Precision-recall curve of RF classifier for Seyfert galaxies (top panel) and quasars (bottom panel), respectively. The thresholds of 0.5 and 0.8 are also marked with blue circles in each panel.
    }
    \label{fig:pre_rec}
\end{figure}

The top panel of Figure \ref{fig:matrix} shows the confusion matrix of the test set. The number of Seyfert galaxies misclassified as quasars (false quasars, 1758) are similar with that of quasars misclassified as Seyfert galaxies (false Seyfert galaxies, 1216). A classifier defines a source class base on the predicted class probabilities. The default threshold is 0.5, which means if the predicted probability of quasars is higher than 0.5 for an objects, it will be classified as a quasar. Figure~\ref{fig:pre_rec} shows the precision-recall curve of RF classifier for Seyfert galaxies and quasars, respectively. The precision-recall pairs are displayed for different probability thresholds, where the probability thresholds of 0.5 and 0.8 are also marked. It is obvious that precision increases as recall decreases, and vice versa. Increasing the threshold of predicted probability would reduce the number of misclassified sources, thereby improving the precision of both quasars and Seyfert galaxies. However, sources failing to meet the new threshold would be considered as unclassified, which consequently lowers the recall. In the bottom panel of Figure \ref{fig:matrix}, we show the confusion matrix of the test set with the threshold of predicted probability being 0.8. The metrics are recalculated and listed in Table~\ref{tab:performance}. With this new threshold, the precisions for identifying quasars and Seyfert galaxies have improves to 98.7 percent and 94.7 percent, respectively. Increasing the classification threshold could be useful to construct samples of quasar and Seyfert galaxy with high purity. On the other hand, the completeness of Seyfert galaxies decreases to 0.470. In order to obtain a relatively complete sample of Seyfert galaxies, a lower threshold (e.g., smaller than 0.5) is favored.

In Figure~\ref{fig:pca}, the predicted probabilities of quasars derived from the RF classifier are overplotted in the PCA diagram. The regions where quasars in the main sample concentrate show higher probabilities of quasars. Sources with lower probabilities lie along a non-overlapping tail of Seyfert galaxies and quasars in the main sample. Seyfert galaxies in the main sample concentrate in the transition region of the probabilities, where the probabilities range from 0.2 to 0.8. These results confirm the partial separation between quasars and Seyfert galaxies. The sources in the transition regions would be difficult to classify correctly, which is likely due to the continuous distributions from quasars to Seyfert galaxies.  
\begin{table*}{}
\centering
\caption{Performance of supervised classifier. \label{tab:performance}}
\begin{tabular}{ccccccc}
\hline\hline
    \multirow{2}{*}{Algorithm} & \multicolumn{3}{c}{Quasar} & \multicolumn{3}{c}{Seyfert galaxy} \\
     & precision & recall & $F_1$ score & precision & recall & $F_1$ score \\
\hline
    RF     & 0.962 & 0.973 & 0.967 & 0.822 & 0.762 & 0.791 \\
    RF (threshold 0.8)  & 0.987 & 0.919 & 0.952 & 0.947 & 0.470 & 0.628 \\
    RF (excluding magnitude)  & 0.953 & 0.969 & 0.961 & 0.790 & 0.708 & 0.747 \\
    RF (subsample)  & 0.726 & 0.844 & 0.781 & 0.742 & 0.584 & 0.654 \\
    RF+SMOTE   & 0.979 & 0.943 & 0.961 & 0.717 & 0.879 & 0.790 \\
    XGB    & 0.963 & 0.970 & 0.966 & 0.807 & 0.771 & 0.789 \\
    XGB+SMOTE  & 0.985 & 0.926 & 0.955 & 0.670 & 0.914 & 0.773 \\
 \hline
\end{tabular}
\end{table*}

\begin{figure}[ht!]
    \centering
    \includegraphics[width=0.45\textwidth]{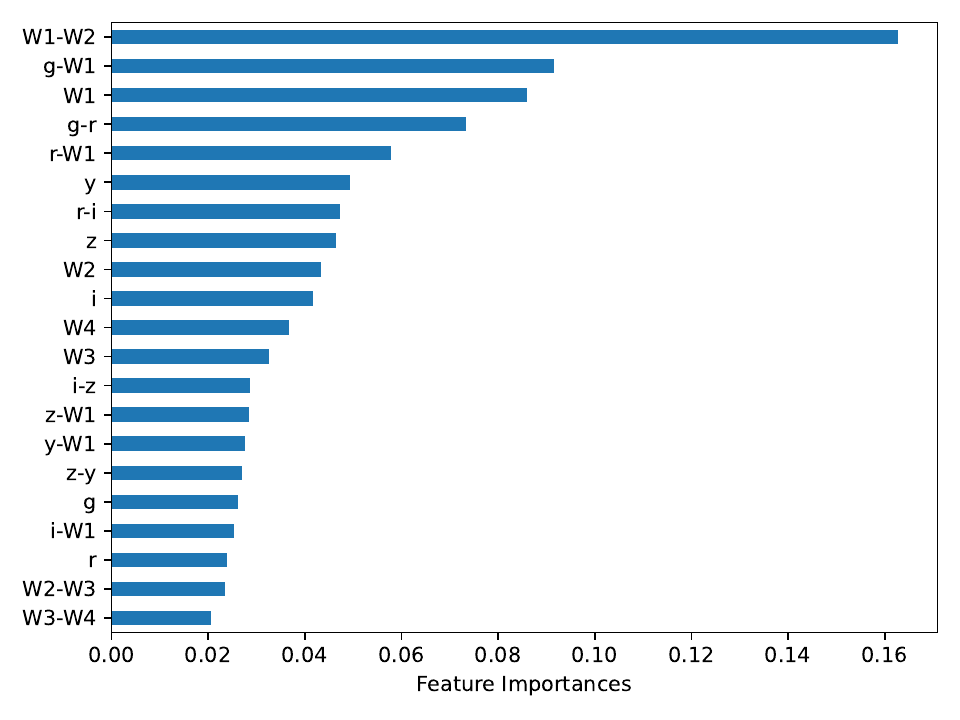}
    \includegraphics[width=0.45\textwidth]{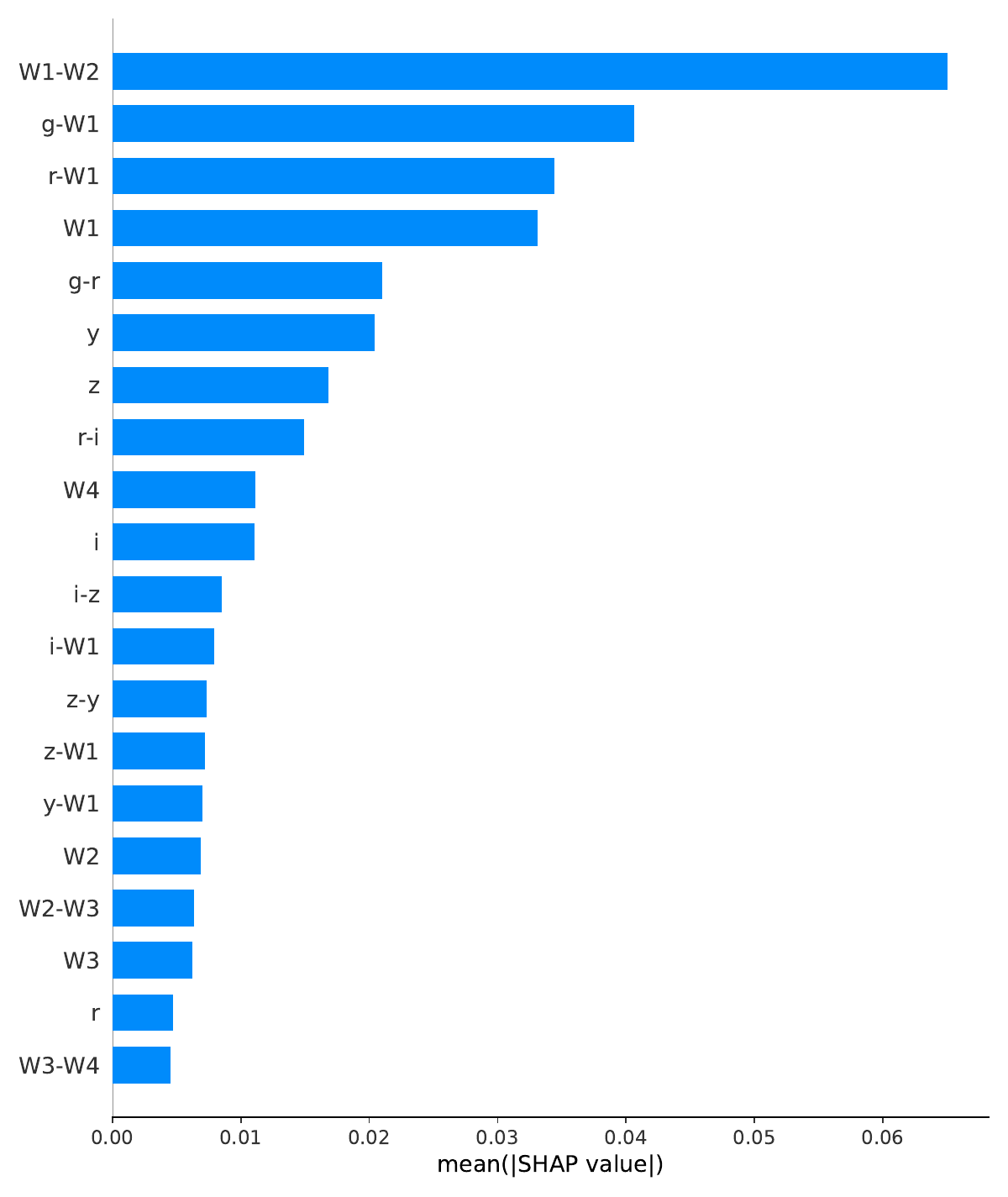}
    \caption{Feature importance for RF classifier. Top panel is derived with the mean decrease in impurity method, while the bottom panel is derived with SHAP method.
    }
    \label{fig:fea_importance}
\end{figure}

RF can evaluate the feature importance by calculating the mean decrease in impurity. If a feature can significantly reduce impurity during segmentation, it would contribute high importance\citep{2014arXiv1407.7502L}. In the top panel of Figure~\ref{fig:fea_importance}, the importance ranking of RF method is plotted. MIR color W1-W2 show the highest importance among the 21 training features, which is consistent with the indication from Figure~\ref{fig:color}. Except W1-W2, g-W1, W1 and g-r are also important to distinguish quasars and Seyfert galaxies. These features are also found to be important in the task of identifying quasars from galaxies or stars~\citep{2019AJ....157....9B, 2020A&A...633A.154L, 2024ApJS..271...54F}. As RF feature importance is known to be biased in the presence of correlated variables and continuous features~\citep{2023arXiv230701932A}, the SHapley Additive exPlanations (SHAP) algorithm is applied to compare with the RF feature importance~\citep{NIPS2017_8a20a862, 2020arXiv190504610L, 2025MNRAS.537..876A}. SHAP is a game theoretic approach to explain the feature’s contribution of a machine learning model. In the bottom panel of Figure \ref{fig:fea_importance}, the importance ranking of SHAP method is plotted. W1-W2 is again the most important feature. The ranking for other features also show little difference with RF method. For the top five important features, four are related to W1 band magnitude in both methods. This might suggest W1 as one of the most important feature to distinguish Seyfert galaxies from quasars. Nevertheless, it remains possible that the intrinsic physical importance is hidden due to these correlated features.

\begin{figure}[htbp]
    \centering
    \includegraphics[scale=0.45]{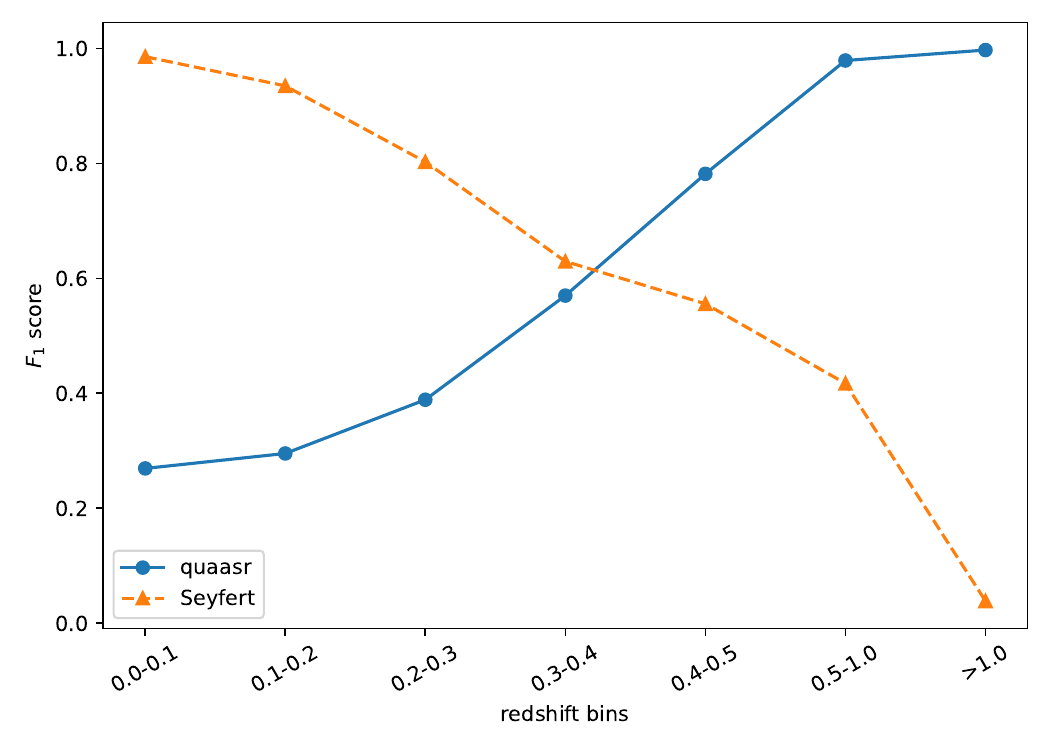}
    \caption{F$_1$ score of the test set at different redshift bins. The blue circles show the variations of F$_1$ score for quasars, while the orange triangles show those for Seyfert galaxies.
    }
    \label{fig:f1_per_z}
\end{figure}

Two further tests are performed to examine whether the RF classifier is sensitive to the combinations of features, or the distributions of redshift and absolute magnitude. The performances of the test set for both tests are also summarized in Table~\ref{tab:performance}. The first test uses a different combination of feature excluding magnitude measurements (i.e., W1-W2, W2-W3, W3-W4, g-r, r-i, i-z, z-y, g-w1, r-W1, i-W1, z-W1, y-W1). The performance gets just slightly worse than the best classifier. The second test is performed with a subsample, which considers the sources within a narrow range of both redshift (0.4 to 0.5) and B band absolute magnitude (-26 to -19). The subsample contains 3294 Seyfert galaxies and 4380 quasars. The performance becomes worse for both quasars and Seyfert galaxies, although these two types of sources could still be distinguishable. In Figure~\ref{fig:f1_per_z}, the overall behavior where F$_1$ score varying with redshift is plotted. Quasar and Seyfert galaxies show inverse behaviors. The performance of quasars gets better with increasing redshift, while it is opposite for Seyfert galaxies. This trend may partly arise from the class imbalance between Seyfert galaxies and quasars in the training sample, where the fraction of Seyfert galaxies decreases with redshift (see Section~\ref{sec:Application} and Figure~\ref{fig:count_ratio}).

\subsection{Applying to DESI DR1} \label{sec:Application}
As a classifier based on photometric data from all-sky surveys, our model can be easily applied to distinguish quasars and Seyfert galaxies for unclassified surveys. As mentioned in Section~\ref{sec:intro}, quasar and Seyfert galaxies can also be separated based on the optical spectroscopic observation. In order to evaluate the reliability and generalization of our classifier, we compare the classification results of our classifier with those from spectroscopic surveys. SDSS and DESI are two of the largest optical spectroscopic surveys currently~\citep{2000AJ....120.1579Y, 2016arXiv161100036D, 2016arXiv161100037D}. As the quasar and AGN samples in Milliquas catalog are collected mainly from SDSS~\citep{2021arXiv210512985F, 2023OJAp....6E..49F}, DESI DR1\footnote{https://data.desi.lbl.gov/doc/releases/dr1/vac/agnqso/} (\citealt{2025arXiv250314745D}, Juneau et al. in preparation) is used to construct a prediction sample. The AGN/QSO Value-Added Catalog of DESI DR1 labels sources with a \texttt{SPECTYPE}, which includes \texttt{`QSO'}, \texttt{`GALAXY'}, and \texttt{`STAR'} according to the results of Redrock fitting (Bailey et al., in prep.). \texttt{AGN\_MASKBITS} labels whether a source satisfy the detailed AGN selection criteria. \texttt{`BROAD\_LINE'} is required for type 1 AGNs. Sources with \texttt{SPECTYPE} = \texttt{`QSO'} and \texttt{AGN\_MASKBITS} = \texttt{`BROAD\_LINE'} are treated as type 1 quasars. \texttt{AGN\_MASKBITS} = \texttt{`BPT\_ANY\_SY'} means a source is classified as Seyfert galaxies based on BPT diagnostic. It will be treated as a type 1 Seyfert galaxy when \texttt{SPECTYPE} = \texttt{`GALAXY'} and \texttt{AGN\_MASKBITS} = \texttt{`BROAD\_LINE'} are also satisfied. The AGN/QSO Value-Added Catalog of DESI DR1 is cross-matched with AIIWISE and PS1 with 1\arcsec~matching radius. After removing the sources with multiple matches and null magnitude measurements, there are 715,400 sources left (hereafter prediction sample). The prediction sample contains 135,431 quasars, 5,642 Seyfert galaxies, and 574,327 sources of other types.

\begin{figure*}[htbp]
    \centering
    \includegraphics[scale=0.85]{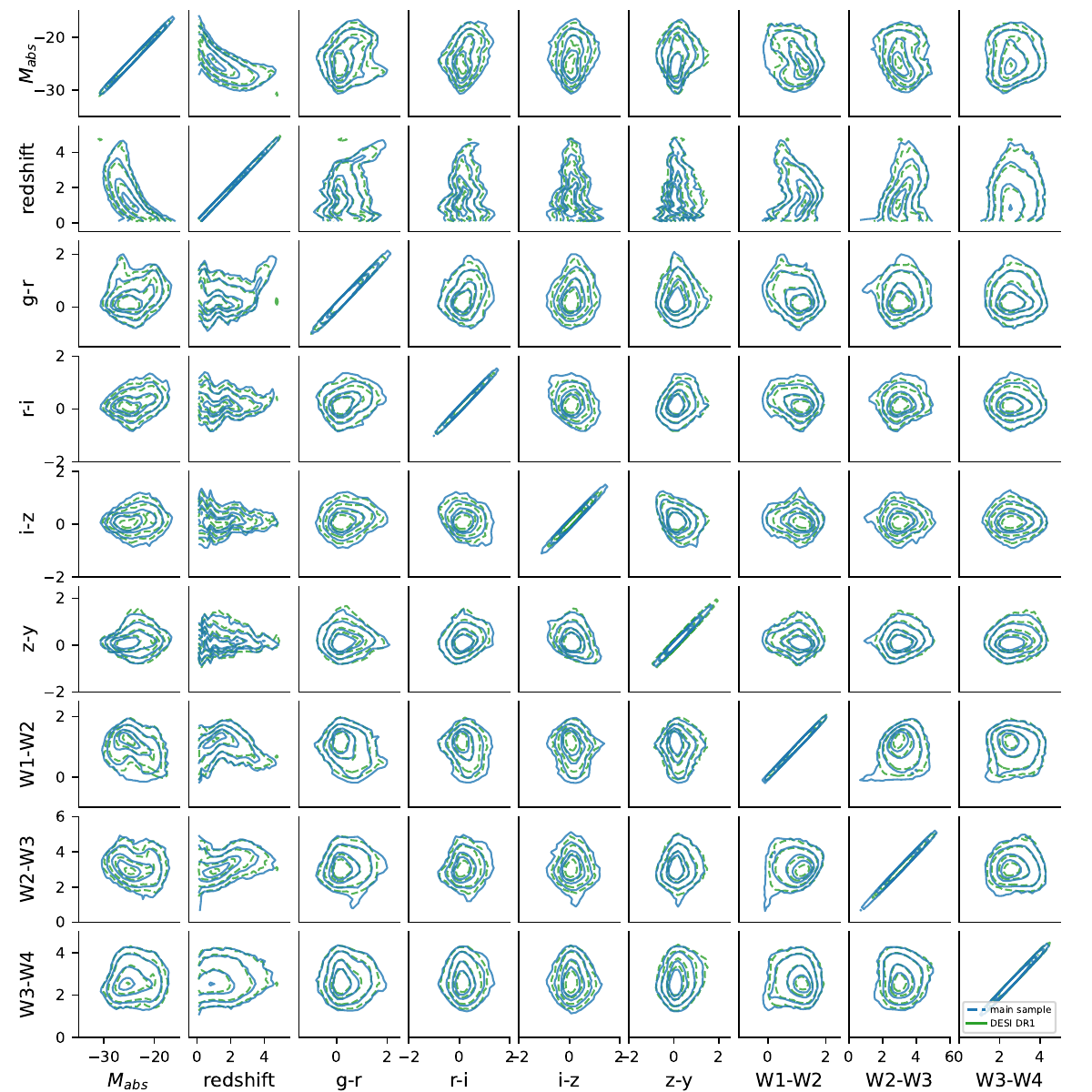}
    \caption{Distributions of features between the main sample (blue contours) and the prediction sample (green contours). The levels of contours are [10, 100, 1000, 3000]. 
    }
    \label{fig:color_comp}
\end{figure*}

The distributions of color indices, absolute magnitude, and redshift are compared for quasars and Seyfert galaxies in the main sample and the prediction sample (Figure~\ref{fig:color_comp}). No obvious offsets are visible in the two-dimensional distributions, except for the low density tails in the main sample for W1-W2 and W2-W3. K-S tests yield D statistics ranging from 0.01 (W2-W3) to 0.08 (g-W1). Given the extremely large sample sizes, the corresponding p-values approach zero ($ < 10^{-8}$ for W2-W3). The small D statistics indicate that the differences in the distributions of these two samples are minimal.

\begin{figure}[htbp]
    \centering
    \includegraphics[scale=0.45]{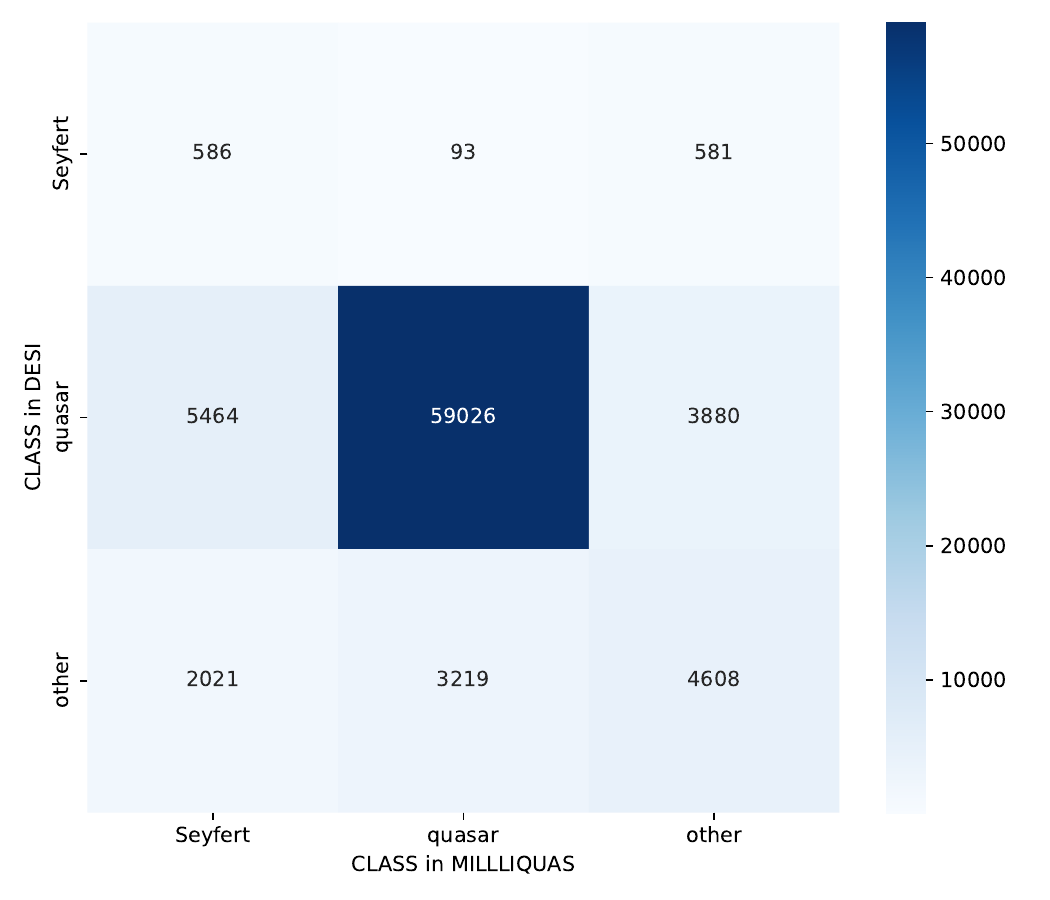}
    \caption{Agreement matrix for the common sources between the prediction sample and Milliquas.}
    
    \label{fig:matrix_comp}
\end{figure}

There are 79478 common sources between the prediction sample and Milliquas. The classifications in two samples are then compared. Figure~\ref{fig:matrix_comp} shows the agreement matrix of them. The majority of quasars are matched in both samples. The count of Seyfert galaxies classified by DESI (1260) is much less than that classified by Milliquas (8071), while most Seyfert galaxies in Milliquas are classified as quasars in DESI (5464). There are 586 matched Seyfert galaxies between two samples. 

The RF classifier is applied to the prediction sample. Table \ref{tab:prediction} shows the confusion matrix of the prediction sample, where the classifications of DESI DR1 are treated as the reference labels. For 135,431 DESI quasars, 119,052 ones are correctly classified, while the number of true Seyfert galaxies are 4,048. Other-type sources are also classified as type 1 AGNs by our classifier. 213,365 sources are classified as quasars, while the other 360,962 are classified as Seyfert galaxies. For a valid classifier, we expect these sources of other types could have lower predicted probability. This means if we increase the threshold, other-type sources would change to be unclassified. When the classification threshold is set to 0.8, 349,880 sources are classified. The fraction of misclassification reduces from 82.8 percent for the default threshold to 67.6 percent. The number of true quasars (111,032) is larger than the number that others are misclassified as quasars (74,147), which indicates that improving the threshold is helpful to identify quasars with high significance. However, true Seyfert galaxies is rare. The majority of Seyfert galaxies classified by our classifier are not type 1 AGNs, although the fraction of correct classified Seyfert galaxies also increases with threshold increasing. When the classification threshold is set to 0.99, the fraction of misclassification can reduce to 27.6 percent. The new confuse matrixes are also listed in Table \ref{tab:new_threshold}.

\begin{table}
\centering
\tiny
\caption{Confusion matrix of prediction sample}
\label{tab:prediction}
\begin{tabular}{ccccc}
\hline\hline   
    & & \multicolumn{3}{c}{Predicted Labels} \\
    &  & quasar & Seyfert & recall \\
\hline
     \multirow{4}{*}{Reference Labels} & quasar & 119052 & 16379 (1414)$^{b}$  & 0.879 (0.988) \\
     & Seyfert & 1594 & 4048 (19013) & 0.717 (0.923) \\
     & other & 213365 & 360962 & -- \\
     & precision$^{a}$ & 0.987 & 0.198 (0.931) &  \\
\hline
\end{tabular}
\tablefoot{\\$^{a}$ The numbers of others are ignored when calculating the precision.\\
     $^{b}$ In 16,379 false Seyfert galaxies, the majority of them have g band absolute magnitude fainter than $-22.3$, which could be true Seyfert galaxies. The precision and recall are recalculated based on this assumption. The results are listed in the brackets.}
\end{table}

\begin{table}
\centering
\fontsize{7}{5.0}\selectfont
\caption{Confusion matrix of prediction sample when the classification threshold is set to 0.8 and 0.99}
\label{tab:new_threshold}
\begin{tabular}{cccc}
\hline\hline   
    & & \multicolumn{2}{c}{Predicted Labels} \\
    &  & quasar & Seyfert \\
\hline
     \multirow{4}{*}{Reference Labels} & quasar & 111032/83692$^{b}$ & 7897/1183 \\
     & Seyfert & 445/20 & 2244/638 \\
     & other & 74147/8224 & 154115/22757 \\
     & fraction of misclassification$^{a}$ & \multicolumn{2}{c}{0.676/0.276} \\
\hline
\end{tabular}
\tablefoot{\\$^{a}$ The fraction is calculated by 1 - (True quasars + True Seyfert galaxies)/Total sources.\\
     $^{c}$ The numbers before and after the slashes represent the source counts when the classification threshold is 0.8 and 0.99, respectively.}
\end{table}

\begin{figure}[htbp]
    \centering
    \includegraphics[scale=0.45]{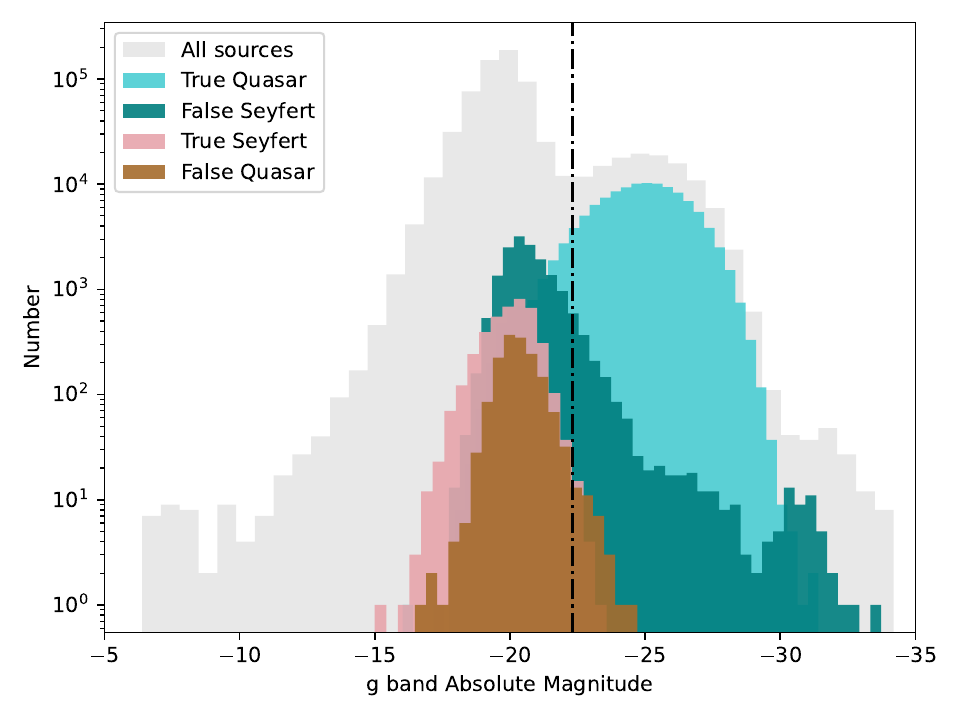}
    \includegraphics[scale=0.45]{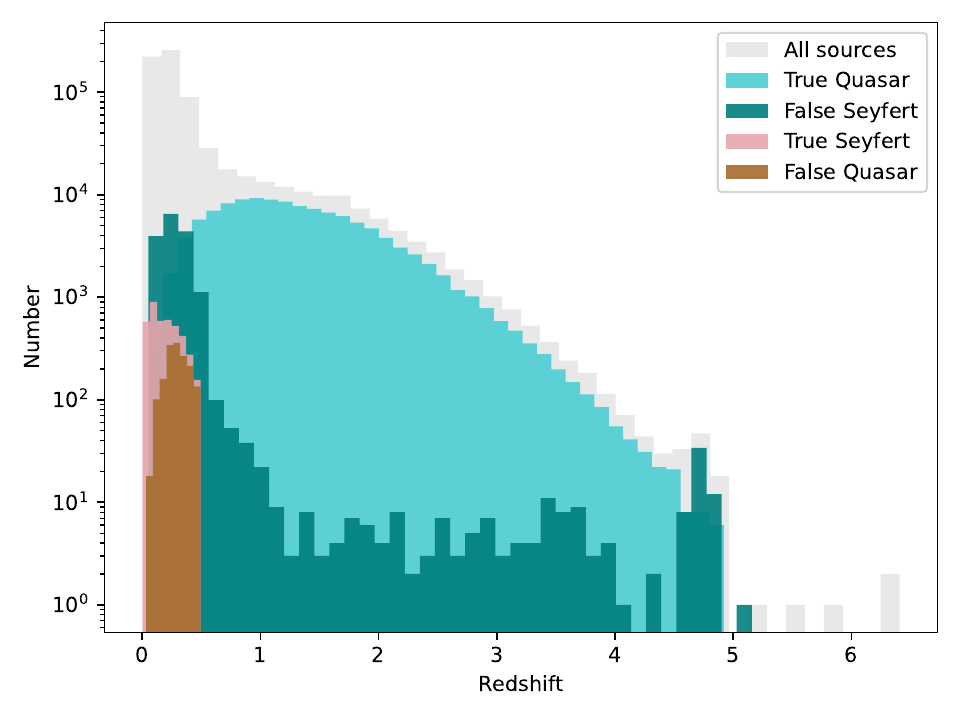}
    \caption{The distributions of g band absolute magnitude and redshift for the prediction sample. The gray bars represent the whole sample in the prediction sample. The cyan and pink bars represent the quasars and Seyfert galaxies in DESI DR1, respectively, who are correctly classified by our classifier. The green bars represent quasars in DESI DR1 misclassified as Seyfert galaxies by our classifier, while the brown bars represent Seyfert galaxies misclassified as quasars. The dot-dashed line in the top panel represent the critical absolute magnitude to separate quasars and Seyfert galaxies.
    }
    \label{fig:pred_pro}
\end{figure}

The evaluation metrics can also be calculated for the prediction sample, when the number of other-type sources are ignored. The precision and recall of quasars are 0.987 and 0.879, respectively. The metrics for Seyfert galaxies are much worse than those of quasars, which are precision $= 0.198$ and recall $= 0.717$, respectively\footnote{The performance is similar when the 79478 common sources in the training sample are excluded from the prediction sample, which are 0.975/0.842 and 0.217/0.670 for precision/recall of quasars and Seyfert galaxies, respectively.}. The poor performance of Seyfert galaxies can be caused by the difference in the feature distributions between the training and prediction sample (Figure~\ref{fig:color} and~\ref{fig:color_comp}). To clarify this, we first explore the distributions of absolute magnitude and redshift for the prediction sample, which are shown in Figure~\ref{fig:pred_pro}. The green and orange shaded areas show the false Seyfert galaxies and false quasars, respectively. Majority of the false Seyfert galaxies show g band absolute magnitude larger than $-22.3$\footnote{The g band divider between quasars and Seyfert galaxie is not defined in the literature. Transformations from g band to B band magnitude for quasars require the estimations of u-g colors~\citep{2005AJ....130..873J}, which is lacking in PS1 DR1. Therefore, we do not attempt to derive a new divider for g band absolute magnitude. Interestingly, $-22.3$ is generally divides the two peaks in the distribution of g band absolute magnitude for all sources in the prediction sample (the gray basrs in the top panel of Figure~\ref{fig:pred_pro}).}, which follow the definition of Seyfert galaxies. Meanwhile, they also locate at low redshift. These sources might be true Seyfert galaxies, while their optical spectra is dominated by the radiation from AGNs. If we classify the 14,965 quasars with absolute magnitude larger than $-22.3$ as Seyfert galaxies, there are only 1,414 DESI quasars misclassified as Seyfert galaxies by our classifier. The model performance on classifying Seyfert galaxies would then improve significantly. The precision and recall are 0.931 and 0.923, respectively. The recall of identifying quasars also improves to 0.988.

\begin{figure}[htbp]
    \centering
    \includegraphics[scale=0.45]{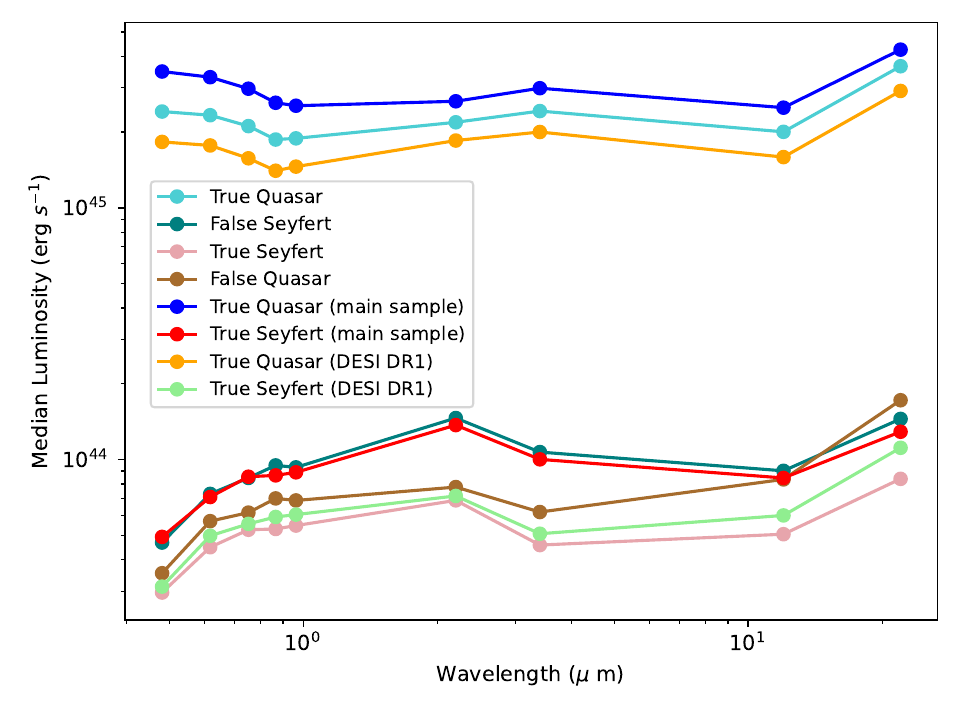}
    \caption{The median SEDs for the main sample and the prediction sample. The cyan, pink, green and brown lines show the true quasars, true Seyfert galaxies, false Seyfert galaxies, and false quasars classified by our classifier. The blue and red lines show the median SEDs of quasars and Seyfert galaxies in the main sample, respectively. The orange and light green lines show the quasars and Seyfert galaxies classified by the spectroscopy method, respectively. 
    }
    \label{fig:mean_sed}
\end{figure}

In order to explore the properties of the misclassified Seyfert galaxies further, we calculate the median SEDs for quasars and Seyfert galaxies. Figure~\ref{fig:mean_sed} shows the median SEDs for the correctly and incorrectly classified quasars and Seyfert galaxies by our classifier, as well as those in the main sample and DESI DR1. The shapes of the median SEDs for true Seyfert galaxies and true quasars identified by our classifier are very similar with those in the main sample, although the luminosities of the former are slightly smaller than those of the later. The false Seyfert galaxies show consistent median SEDs with Seyfert galaxies in the main sample. In addition, the SEDs of quasars and Seyfert galaxies classified by spectroscopic method also show similar shapes with the true quasars and true Seyfert galaxies identified by our classifier, despite mild offsets at W3 and W4 bands for Seyfert galaxies. These results indicate that our classifier can reconstruct the shapes of SEDs for quasars and Seyfert galaxies, and the Seyfert galaxies identified by our classifier should be true Seyfert galaxies. The shape of SED for false quasars is different with all other sources, which indicates a distinct population in type 1 AGNs.

\begin{figure}[htbp]
    \centering
    \includegraphics[scale=0.45]{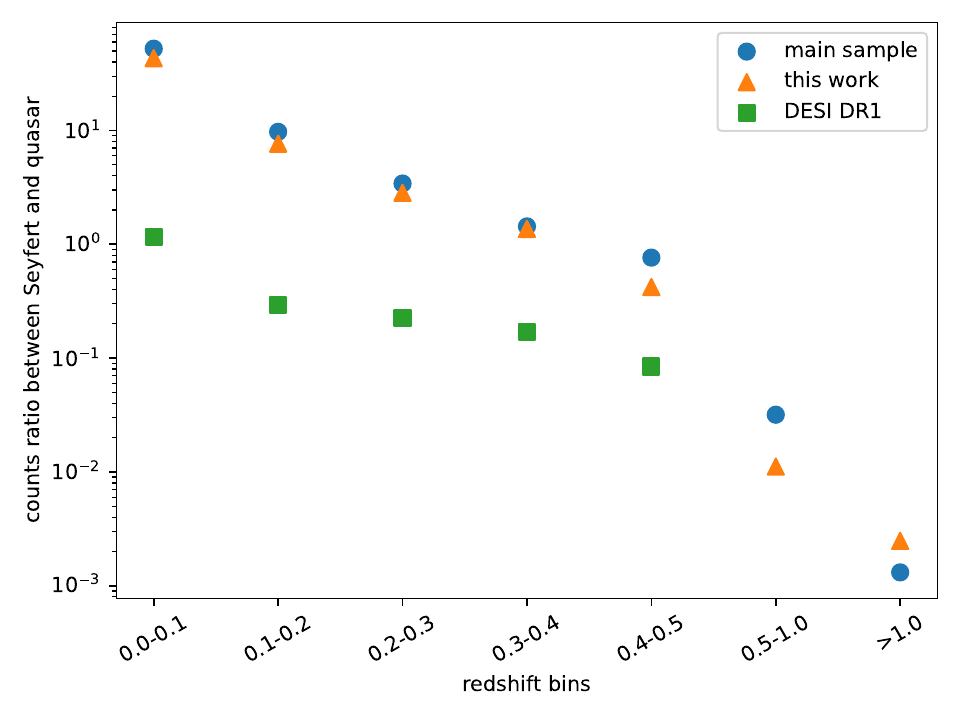}
    \caption{The counts ratio between Seyfert galaxies and quasars at different redshift bins for different classification methods. The blue cycles, orange triangles and green squares represent the counts ratios for the main sample, the classifications of the prediction sample based on our classifier and spectroscopy method, respectively.
    }
    \label{fig:count_ratio}
\end{figure}

In Figure~\ref{fig:count_ratio}, we also compare the counts ratio between Seyfert galaxies and quasars at different redshift bins. Compared with the main sample, the fraction of Seyfert galaxies identified by spectroscopic method are smaller. Especially, the spectroscopic method identifies no Seyfert galaxies at z $>$ 0.5. As a contrast, our classifier finds similar fractions of Seyfert galaxies with the main sample at all redshift bins. It appears that a large fraction of Seyfert galaxies are missed by spectroscopic classification, although it cannot be ruled out that the main sample and our classifier overestimate the fraction of Seyfert galaxies.

\begin{figure}[htbp]
    \centering
    \includegraphics[scale=0.45]{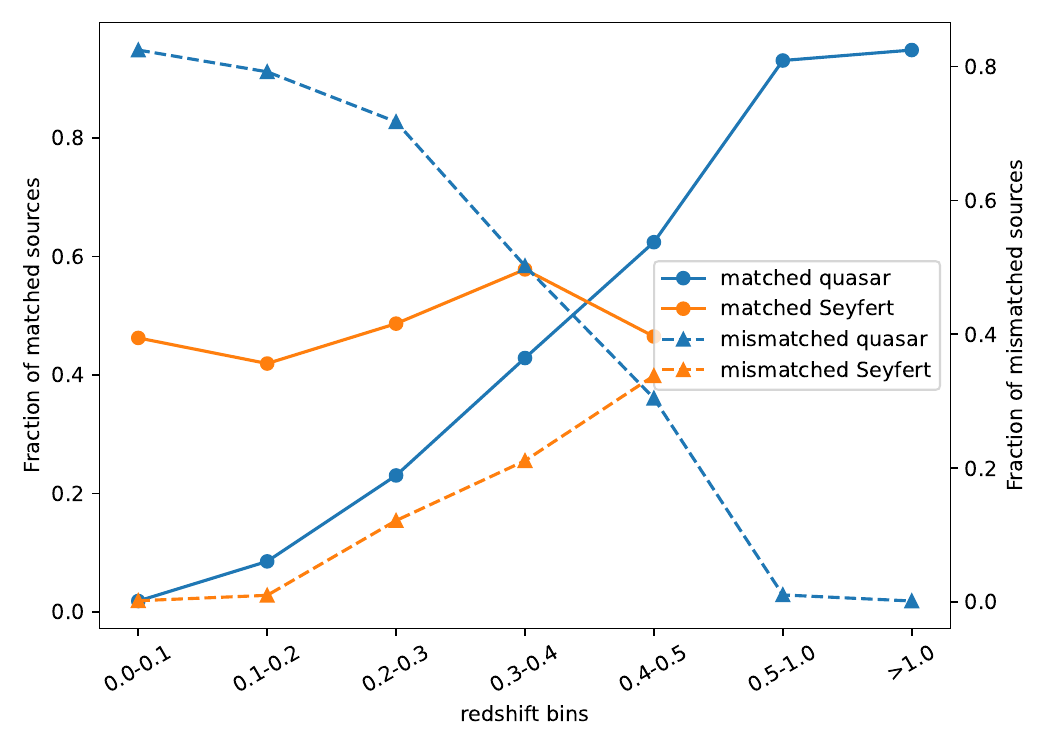}
    \caption{Fraction of matched and mismatched sources between the prediction sample and Milliquas at different redshift bins. The left Y-axis shows the fraction of matched sources, while the right Y-axis shows the fraction of mismatched sources. The blue/orange cycles and blue/orange triangles represent the matched and mismatched fraction for quasars/Seyfert galaxies, respectively.
    }
    \label{fig:matched_per_z}
\end{figure}

We investigate the difference of the counts ratio in detail with the matched sample between the prediction sample and Milliquas. The fractions of matched and mismatched sources between these two samples are compared at different redshift bins. The fractions are defined as the ratio between matched or mismatched quasars/Seyfert galaxies and total quasars/Seyfert galaxies classified by DESI in each redshift bin. Matched sources are those with consistent classifications from two catalogues. Mismatched quasars/Seyfert galaxies are objects labelled as quasars/Seyfert galaxies in DESI, but classified as Seyfert galaxies/quasars by Milliquas. The results are shown in Figure~\ref{fig:matched_per_z}. Mismatched quasars are mainly concentrated at low redshift, with their fraction exceeding 50 percent for redshift less than 0.4. Matched quasars exhibit opposite trend, which become dominant at high redshift. The fractions of matched Seyfert galaxies show little redshift-dependent variation, while mismatched Seyfert galaxies remain a minority at all redshift bins. This provides clear evidence that the difference of the counts ratio between the main sample and prediction sample is caused by the mismatched quasars between the two classification methods.

In Figure~\ref{fig:pred_pro}, the misclassified Seyfert galaxies show peaks at redshift between 4 and 5, and the g band absolute magnitude around $-30$. We explore the properties of these peculiar sources with the interactive spectrum browser of DESI project\footnote{e.g., https://www.legacysurvey.org/viewer/desi-spectrum/dr1/\\targetid39627346113731185}. For all the 54 misclassified Seyfert galaxies with $z > 4$ and $M_g < -27$, the redshift of them are less than 0.2 from the interactive spectrum browser, while the images for most of them show obvious extended structures from host galaxies. Our classifier indeed reveals their Seyfert galaxies properties based only on the photometric data. We note that the high performance of the RF classifier partly reflects the classification criteria between quasars and Seyfert galaxies in the training sample, it should not be regarded as evidence for an intrinsic physical distinction between these two populations.

\section{Summary} \label{sec:summary}
Quasar and Seyfert galaxy are two important subclasses of type 1 AGNs. Clarifying the difference between them is useful to understand the co-evolution of supermassive black hole and their host galaxies. In this paper, we provide a machine learning based classifier to distinguish type 1 quasars and type 1 Seyfert galaxies with photometric data of MIR and optical bands. By cross-matching Milliquas, AllWISE and PS1 DR1, we obtain a sample of 150,381 quasars and 24,881 Seyfert galaxies with optical and MIR features. Although the color-color diagrams shows that quasars and Seyfert galaxies are indistinguishable for most optical/MIR colors, the results of PCA visually confirm that these two types of sources are partially distinguishable, when multi-dimensional features are considered. The features of optical and MIR magnitudes and colors, combined with colors between optical and MIR bands are then trained by RF algorithms to derive the classifier. For quasars, the RF classifier shows good performance with both high precision (0.962) and recall (0.973). The model performance is worse for Seyfert galaxies with precision $= 0.822$ and recall $= 0.762$. Increasing the classification threshold to 0.8 would achieve a high precision (0.947) at the cost of recall (0.470).

To verify the generalization of our classifier, it is applied to predict the classification of sources in Value-Added Catalog of DESI DR1. The results are compared with those from spectroscopic classification. For quasars, our classifier achieve a good performance again. However, a large number of spectroscopically classified quasars are classified as Seyfert galaxies by our classifier. The majority of them show g band absolute magnitude larger than $-22.3$. If these faint quasars in DESI are regarded as Seyfert galaxies, our classifier would also perform well on classifying Seyfert galaxies.

We explore the robustness of our classifier with another two approaches. The median SEDs of Seyfert galaxies identified by our classifier are similar with those of the training sample, as well as the ones classified by spectroscopic method. The counts ratio between Seyfert galaxies and quasar also indicates that Seyfert galaxies could be completely identified by our classifier. These Seyfert galaxies can meet the luminosity threshold of Milliquas, even though they might not satisfy the criteria for spectroscopic classification.

Our classifier can be applied to any photometric surveys at optical and MIR bands, and provide quasars candidates with high precision. Sources who are not type 1 AGNs could be misclassified for a pure photometric catalog. Increasing the classification threshold can, to a large extent, reduce the misclassified sources. For Seyfert galaxies, preliminary information about the broad emission lines can help to reduce the misclassifications. Our classifier can also identify Seyfert galaxies who might be missed by the spectroscopic classification. This would be useful for deriving a relatively complete sample of Seyfert galaxies and exploring the evolution of AGNs and their host galaxies at different redshifts.

\begin{acknowledgements}
We are grateful to the anonymous referee for his/her constructive comments. We thank Zhen-Yi Cai for helpful discussions to improve this manuscript. This research is funded by the National Natural Science Foundation of China (NSFC; grant No. 12003014, 12573077).\\

This publication makes use of data products from Pan-STARRS. The Pan-STARRS1 Surveys (PS1) and the PS1 public science archive have been made possible through contributions by the Institute for Astronomy, the University of Hawaii, the Pan-STARRS Project Office, the Max-Planck Society and its participating institutes, the Max Planck Institute for Astronomy, Heidelberg and the Max Planck Institute for Extraterrestrial Physics, Garching, The Johns Hopkins University, Durham University, the University of Edinburgh, the Queen's University Belfast, the Harvard-Smithsonian Center for Astrophysics, the Las Cumbres Observatory Global Telescope Network Incorporated, the National Central University of Taiwan, the Space Telescope Science Institute, the National Aeronautics and Space Administration under Grant No. NNX08AR22G issued through the Planetary Science Division of the NASA Science Mission Directorate, the National Science Foundation Grant No. AST-1238877, the University of Maryland, Eotvos Lorand University (ELTE), the Los Alamos National Laboratory, and the Gordon and Betty Moore Foundation.\\

This publication makes use of data products from the Wide-field Infrared Survey Explorer, which is a joint project of the University of California, Los Angeles, and the Jet Propulsion Laboratory/California Institute of Technology, funded by the National Aeronautics and Space Administration.\\

This research used data obtained with the Dark Energy Spectroscopic Instrument (DESI). DESI construction and operations is managed by the Lawrence Berkeley National Laboratory. This material is based upon work supported by the U.S. Department of Energy, Office of Science, Office of High-Energy Physics, under Contract No. DE–AC02–05CH11231, and by the National Energy Research Scientific Computing Center, a DOE Office of Science User Facility under the same contract. Additional support for DESI was provided by the U.S. National Science Foundation (NSF), Division of Astronomical Sciences under Contract No. AST-0950945 to the NSF’s National Optical-Infrared Astronomy Research Laboratory; the Science and Technology Facilities Council of the United Kingdom; the Gordon and Betty Moore Foundation; the Heising-Simons Foundation; the French Alternative Energies and Atomic Energy Commission (CEA); the National Council of Humanities, Science and Technology of Mexico (CONAHCYT); the Ministry of Science and Innovation of Spain (MICINN), and by the DESI Member Institutions: www.desi.lbl.gov/collaborating-institutions. The DESI collaboration is honored to be permitted to conduct scientific research on I’oligam Du’ag (Kitt Peak), a mountain with particular significance to the Tohono O’odham Nation. Any opinions, findings, and conclusions or recommendations expressed in this material are those of the author(s) and do not necessarily reflect the views of the U.S. National Science Foundation, the U.S. Department of Energy, or any of the listed funding agencies.\\

This research has made use of the VizieR catalog access tool, CDS, Strasbourg, France~\citet{10.26093/cds/vizier}. The original description of the VizieR service was published in \citet{2000A&AS..143...23O}.
\end{acknowledgements}

\bibliographystyle{aa} 
\bibliography{bib} 

\end{document}